\documentclass[
aip,
 amsmath,amssymb,
 reprint,%
]{revtex4-1}

\usepackage{graphicx}
\usepackage{dcolumn}
\usepackage{bm}

\usepackage[utf8]{inputenc}
\usepackage[T1]{fontenc}
\usepackage{mathptmx}
\usepackage{etoolbox}
\usepackage{siunitx}
\usepackage{caption}
\usepackage{subcaption}

\makeatletter
\def\@email#1#2{%
 \endgroup
 \patchcmd{\titleblock@produce}
  {\frontmatter@RRAPformat}
  {\frontmatter@RRAPformat{\produce@RRAP{*#1\href{mailto:#2}{#2}}}\frontmatter@RRAPformat}
  {}{}
}%
\makeatother
\begin{document}

\preprint{AIP/123-QED}

\title[]{Multiscale modeling framework of a constrained fluid with complex boundaries using twin neural networks}
\author{Peiyuan Gao}
 \email{peiyuan.gao@pnnl.gov}

 \affiliation{ 
Advanced Computing, Mathematics and Data Division, Pacific Northwest National Laboratory, Richland, WA, 99354, USA}
\author{George Em Karniadakis}
\affiliation{ 
Advanced Computing, Mathematics and Data Division, Pacific Northwest National Laboratory, Richland, WA, 99354, USA}
\affiliation{ 
Division of Applied Mathematics, Brown University, Providence, RI, 02912, USA
}%
\author{Panos Stinis}
\affiliation{ 
Advanced Computing, Mathematics and Data Division, Pacific Northwest National Laboratory, Richland, WA, 99354, USA}


\begin{abstract}
Abstract: The properties of constrained fluids have increasingly gained relevance for applications ranging from materials to biology.
In this work, we propose a multiscale model using twin neural networks to investigate the properties of a fluid constrained between solid surfaces with complex shapes.
The atomic scale model and the mesoscale model are connected by the coarse-grained potential which is represented by the first neural network. 
Then we train the second neural network model as a surrogate to predict the velocity profile of the constrained fluid with complex boundary conditions at the mesoscale. 
The effect of complex boundary conditions on the fluid dynamics properties and the accuracy of the neural network model prediction are systematically investigated.
We demonstrate that the neural network-enhanced multiscale framework can connect simulations at atomic scale and mesoscale and reproduce the properties of a constrained fluid at mesoscale. 
This work provides insight into multiscale model development with the aid of machine learning techniques and the developed model can be used for modern nanotechnology applications such as enhanced oil recovery and porous materials design.
\end{abstract}

\maketitle

\section{\label{sec:level1}Introduction}

Computational methods in physics, materials science, and engineering have been developed
to describe physical and chemical phenomena in nature. 
In traditional computational methods, the main objective is to find approximate solutions by solving various governing equations
that come from either discrete or continuous models of the problem.
Multiscale modeling has been developed for decades with the aim of providing quantitative tools for assessing key variables of realistic systems.
Multiscale modeling combines models defined at fundamentally different length and time scales within the same overall spatial and temporal domain.
It enables probing phenomena at a smaller scale
 and embedding the relevant mechanisms or transferring the key parameters to a larger scale and predicting the dynamics of the overall system.
In recent years, there has been a tremendous growth of activity regarding multiscale modeling methods, which have been adopted in various fields such as nanomedicine, materials design,
 and fluid mechanics.\cite{horstemeyer2010multiscale,You373,Biferale2007,Silva2009}

Generally, multiscale models can be classified into two types: sequential models and concurrent
models. \cite{Bishara372} In a sequential multiscale model, the models at different scales are connected implicitly.
A high-resolution model, e.g., at the molecular scale, is used to generate necessary
parameters for a model at a larger scale. For instance, the degradation kinetics of an organic species can be computationally derived at the molecular level. The reaction rate constant is then transferred to a model at the organelle scale, enabling study of biochemical processes from that vantage point.
This method is dominant in the applications of multiscale modeling. Some information may be lost when passing the parameters from one scale to another, thus increasing the uncertainty and affecting the accuracy of the prediction at a large scale.
Therefore, properly designing these key parameters and compensating for the loss of information are both very important for building an effective sequential multiscale model.

In a concurrent multiscale model, the quantities from both 
macro- and microscales are computed simultaneously. For example, in the case of an open system of a biomolecule in water, the multiscale model can be built with different resolutions in space. The biomolecule and the water molecules near it are modeled at the atomic scale. The water molecules that are further away
from the biomolecule can be represented by a coarse-grained model, while the distant water molecules are represented through a continuum model. This creates "interfaces" in the spatial domain of the simulated system.\cite{Poblete2010,Potestio2013,Kreis2016} When a molecule crosses such an "interface," its resolution needs to be changed accordingly.\cite{Espanol2015,Kubinco2021} The concurrent model can capture the behavior of the system at different scales simultaneously. However, there is an associated increase in the simulation cost compared to the sequential multiscale model. 

Over the last decade, machine learning (ML) has been increasingly recognized as a promising technology in computational sciences and engineering.\cite{Alber371,Peng2021}
Rapid advances in ML algorithms and data science have advanced computational science into the big data era. 
Atomistic simulations of large systems such as a kind of bulk solid material or a fluid system with first-principles simulation approaches like density functional theory (DFT) are very expensive. As a result, there has been a significant amount of research on building surrogate models of the interaction potential energy surface 
between atoms using ML methods like kernel-based Gaussian approximation potential and neural network (NN) potential,\cite{Deringer382, Bartok2015, Keith2021,Behler2007,Unke2021,Friederich2021}
 which have achieved good accuracy and improved the simulation speed compared with the DFT calculation. 
Also, ML has been introduced into quantum mechanics/molecular mechanics (QM/MM)-like schemes that enable improved multiscale simulations.\cite{Zhang2019}
In addition, several works have used ML for constructing coarse-grained potentials for biomolecular systems through the matching of mean forces.\cite{wang2019,Wangj2020,John2017,Gkeka2020,Sivaraman2022,Ge2023}
However, most of these works focus on the structural property of individual molecules such as the molecular conformation. There are few investigations on the bulk properties, which are more important for material applications at the macroscopic level. 

It is well known that fluids can exhibit different behaviors when they are confined at the microscopic level and that the continuum equations may therefore not apply.\cite{Chen2014}
The emergence of micro/nanofluidics, primarily triggered by the phenomenal advancements in generating small-scale geometrical features through micro- and nano-fabrication technologies, has simultaneously rekindled interest in several classical areas of fluid dynamics. 
With important advances in nanomaterials science, microfluidics and nanofluidics have provided better understanding of many physical phenomena such as lubrication, adhesion, adsorption, and wetting, \cite{Liu2012,Darhuber2005,Adera2020} 
as well as significant insights into various applications.\cite{Arzani2022,Hou2022} Microfluidic/nanofluidic transport is often characterized by sharply demarcating variations in the fluid properties over the interfacial regions as compared to those in
the bulk. As the confinement size becomes narrower, the shape of the boundary and interactions between the wall
and fluid tend to play a more prominent and decisive role.\cite{Chakraborty2010}
The study of thermodynamical and dynamical properties of confined fluids by surfaces of different morphologies, especially at the micro-nanometer scale, is very important for many engineering applications. 
However, it is still difficult to directly measure the properties of constrained micro/nanofluids even with modern experimental techniques such as atomic force microscopy or the surface forces apparatus.  

Molecular simulation approaches such as Monte Carlo (MC) and molecular dynamics (MD) have been used to investigate the structure, thermodynamics, and dynamics in micro/nanofluids systems.\cite{Grandis2006,Puibasset2007,Eslami2008}
The atomistic simulation of constrained fluids remains computationally expensive, especially when one is interested in modeling larger systems with complex boundaries. \cite{Mozaffari2019,Mozaffari2017}
An alternative approach is to use mesoscopic, coarse-grained (CG) models for incorporating particles that are large enough for external (hydrodynamic) forces to influence their dynamics but not so large that thermal (Brownian) forces are negligible.\cite{Guo2016,Guo22016,Ge2023} Dissipative particle dynamics (DPD) is one of those mesoscopic techniques that has been shown to be successful in predicting the dynamic properties of soft matter materials at equilibrium and nonequilibrium conditions.\cite{Santo2021}
The DPD thermostat is momentum-conserving and satisfies the Galilean invariance that is required to follow the Navier–Stokes behavior in the macroscopic limit. 
However, a systematic way to develop DPD models from atomistic scales simulations is still needed, especially for constrained fluid systems.\cite{Flekk1999,Flekk2000,ye2023}
In most previous works, simulations of constrained fluids were carried out only with smooth walls and often focused on the phase transition or the morphology of pure fluid or liquid-liquid mixtures at equilibrium state.\cite{LiuH2012,Gidituri2017} 
Almost all implementations of boundary conditions attempted thus far were meant
to achieve a uniform density distribution or velocity profile close to the analytical solution. Special treatments have been proposed to implement a no-slip boundary
condition in DPD fluid simulations, such as using the frozen wall of particles, collision reflection algorithms, and other methods.\cite{xu2009, BERKENBOS2008, KASITEROPOULOU2011,Thalakkotto2016} Revenga \emph{et al.} used effective forces to represent
the effect of a wall on fluid particles instead of using wall particles.\cite{Revenga1999} For a planar wall, the effective forces can be obtained analytically. But these forces are not sufficient to prevent fluid particles from crossing the wall. When particles
cross the wall, a wall reflection is used to reflect particles back to the fluid. Willemsen \emph{et al.} added an extra image layer of
particles outside the simulation domain.\cite{Willemsen2000} The position and velocity of particles in this layer are determined by the particles inside the simulation domain near the wall, such that the mean velocity of a pair of particles inside and outside the
wall satisfies the given boundary conditions. Wang \emph{et al.} combined the image and frozen particle methods and adjusted them to match the density and velocity profiles.\cite{Wang427} 
Fan \emph{et al.} used frozen particles to represent the wall but added a thin layer to hold the no-slip boundary
and the reflection boundary condition. \cite{Fan2003} The interaction parameter between the particles in the thin layer and the bulk region can be tuned to minimize the distortion of the density profile.
Pivkin and Karniadakis developed the adaptive boundary condition method using discrete wall-fluid forces
to target the density profile obtained from MD simulations. \cite{Pivkin2006,Pivkin2005}
From a microscopic perspective, the solidification phenomena and slip velocity are caused by solid-liquid interactions .\cite{Chen20141}Therefore, it may not be the same as the
analytical solutions, especially for complex boundary conditions.\cite{Li429,Cross2018,Dewangan2022}

On the other hand, ML has been widely applied in fluid mechanics research at a fixed scale. Several NN architectures such as convolutional neural networks (CNNs), 
graph neural networks (GNNs), and graph convolutional neural networks (GCNNs) have been employed to train ML models
 with computational fluid dynamics (CFD) calculation data. These ML models have been used for predicting flow dynamics, optimization, and control.\cite{Brunton2020,Kreyenberg2019,Cais2021,Leverant2021,jin2018,Chen425}
ML and multiscale modeling are capable of complementing each other, resulting in the development of highly reliable and predictive models that effectively explore vast parameter spaces. 
For instance, Lino \emph{et al.} developed GNN models for extrapolating the time evolution of a fluid flow.\cite{Lino2022}
In GNN-based models, previous states are processed through multiple coarsenings of the graph, which enables faster information propagation through the network and improves the capture and forecast of the system state, 
especially for phenomena spanning a range of time scales. The models show good accuracy compared to traditional numerical methods. 
However, it is difficult to solve a problem at different spatial scales using the same ML model because extrapolation is usually problematic. 
For example, an ML model that was trained with macroscale data only works on the macroscopic scale.
Moreover, in the field of fluid mechanics, the physics that connect the macroscale and the micro/nanoscale is not very clear, which makes it difficult to integrate physics into the ML model. 
How to build the multiscale model or multiscale simulation framework with ML methods remains a challenge.

In this work, we propose a multiscale simulation framework incorporating twin NNs to establish a connection between the microscale and the mesoscale. Petroleum, specifically its main component,
octane liquid, was selected as an example. We started with simulation data at the atomistic level, from which we built a CG model using the first NN model as a surrogate for the CG interaction potential. 
The CG interaction potential was integrated 
into a DPD framework, where the atomistic data were also used to estimate the rest of the needed parameters about dynamical properties (dissipation and noise). Through DPD simulations, we generated fluid dynamics data of constrained octane fluid under shear with complex boundary conditions.
These data were used to train the second NN, which is capable of predicting the velocity profile of the fluid with complex boundary conditions under shear, thus enabling a transformation from Lagrangian to Eulerian descriptions for the fluid.  
In this way, the twin NN framework successfully connects the molecular scale and the mesoscale. The framework is expected to extend to larger scales too if one uses more NNs. 

The paper is organized as follows. In Section II, we introduce 
the framework, the methodology details of the multiscale model with NN, and each of the twin NNs. Section III contains results for each NN model, their connection, and discussion. Section IV offers conclusions and suggestions for future work.  

\section{\label{sec:level2}Methodology}
\subsection{\label{sec:level21}Total architecture}
\begin{figure*}
    \centering
    \includegraphics[width=6in]{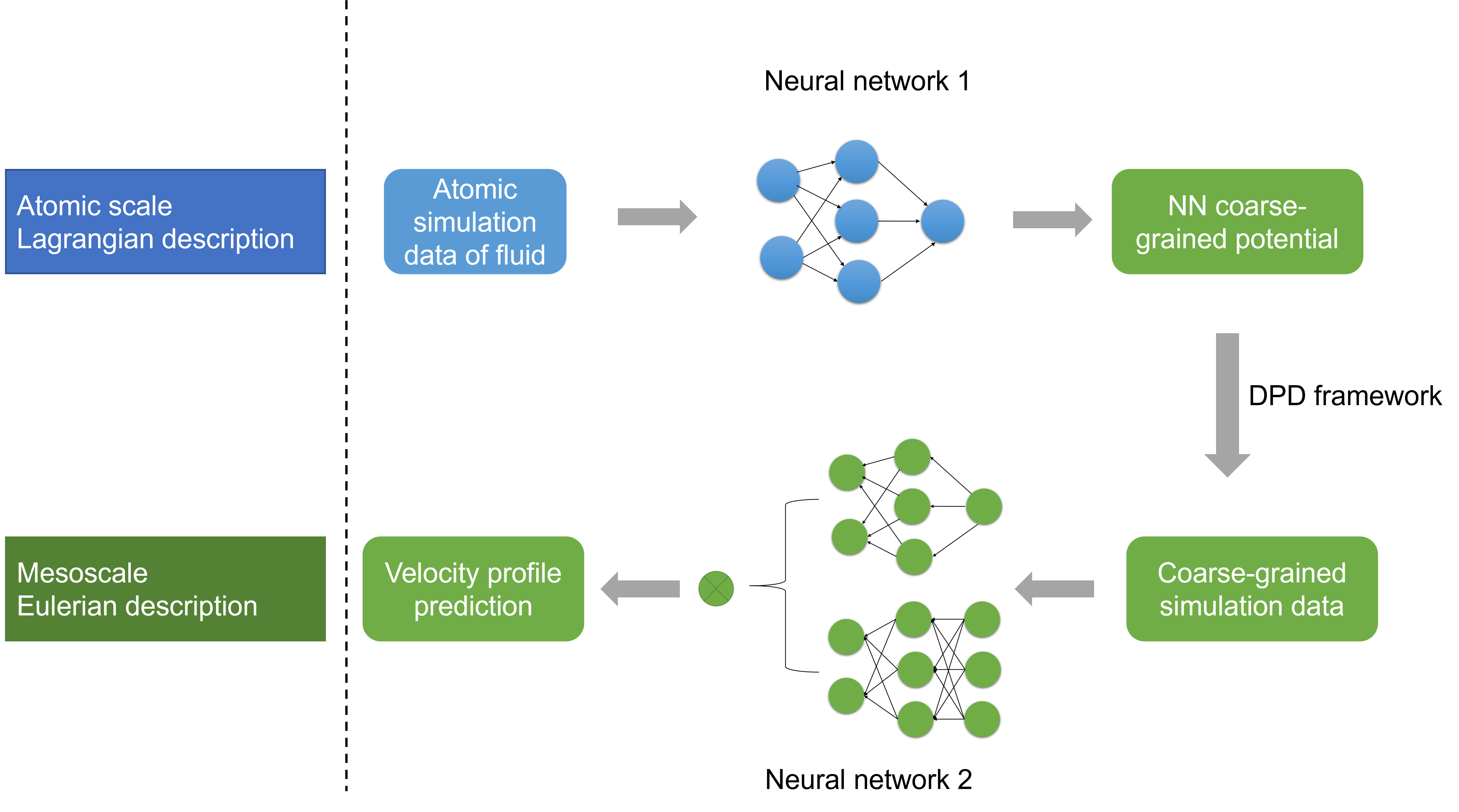}
    \caption{\textsf{The structure of the total framework architecture.}}
    \label{fig:totalarchitecture}
\end{figure*}

In the present work, we built a multiscale modeling framework of fluids constrained by complex boundaries by combining two neural networks. 
As shown in Figure \ref{fig:totalarchitecture}, we first performed atomistic MD simulations of octane and generated trajectory data including the position of atoms, energies, and forces. Then, these data were used as training data for the first NN to obtain a representation of the interaction potential of octane at the CG level.
 The NN CG potential was integrated into a DPD framework.
 With the aid of the DPD framework, the CG model of octane can better reproduce dynamical properties like the diffusion coefficient and viscosity of the fluid in a mesoscale simulation.
Using the DPD framework, we built a model for constrained octane fluid with complex boundary conditions and calculated the velocity data of octane fluid. 
These velocity data and the boundary condition parameters were fed to the second NN model, which consists of a deep neural operator architecture called DeepONet. \cite{Lu2021} The second NN model was trained to predict the velocity profile at mesoscale. 
More details about the construction are provided in the following sections.

\subsection{\label{sec:level22}Atomistic MD simulation}
Our framework is initiated at the atomistic level. The training and validation data for the first neural network consisted of energies, forces, and atomistic positions from a series of MD 
simulations. To prepare the training and validation data, atomistic MD simulations were performed using the Large-scale Atomic/Molecular Massively Parallel Simulator (LAMMPS) \cite{thompson2022lammps}. 
The simulated system is bulk octane, which includes 480 octane molecules. 
The interaction potentials between atoms were represented by the Condensed-phase Optimized Molecular Potentials for Atomistic Simulation Studies (COMPASS) force field\cite{Sun1998}, which is a second-generation force field. 
The COMPASS force field has been validated through prediction of thermodynamic properties of small molecules.\cite{Burrows2021}
The cutoff distances for the Lennard–Jones potential and Coulombic interactions were set to 12 angstrom. A long-range van der Waals (vdW) interaction correction was added. Periodic boundary conditions were applied in three spatial directions.
 The Particle–Particle–Particle–Mesh algorithm (PPPM) was used to calculate the long-range electrostatic interactions.\cite{hockney2021computer,POLLOCK1996}
  Initially, all the molecules were placed randomly in the simulation box.
The initial density was set to its experimental density 0.7 g/cm$^3$.
Energy minimization was performed with the conjugate gradient algorithm to eliminate the high-energy conformation of molecules and to relax the system.
The energy tolerance was set to $10^{-4}$ kcal/mol, and the force tolerance was set to $10^{-6}$ kcal/mol/angstrom. The simulation strategy is a two-step approach, i.e., a pre-equilibrium simulation step and a production simulation step.
In the first step, the system was pre-equilibrated for 5 ns at 298 K and 1 atm in the isothermal-isobaric (NPT) ensemble. The V-rescale thermostat \cite{Bussi2007} and Berendsen barostat \cite{Berendsen1984} were used to keep the temperature and pressure constant.
We confirmed that the MD simulation result with the COMPASS force field can reproduce the experimental density.
 In the second step, 100 production simulations were performed at 298 K in the canonical (NVT) ensemble. The time interval covered by each simulation is 5 ns.
 The temperature of simulation systems was controlled using the Nosé-Hoover thermostat.\cite{nose1984,Martyna1992} 

 The equations of motion were integrated using the velocity Verlet algorithm with a time step of 1 fs. The trajectory files were saved every 500 steps.
Additional 200 ns simulation data at 500 K (40 simulations) were generated to improve the sampling efficiency in the parameter space.
In these simulations, all the bond lengths and angles were fixed because we wanted to explore the parameter space of molecular conformation, which is mostly related to the dihedral angles in a molecule.
Only the coordinates of the atoms were saved during the simulations. The energy and the force of atoms were recalculated at 298 K based on the position of the atoms.

\subsection{\label{sec:level23}The first neural network}
The first neural network (NN1) will be used to represent the interaction potential for the octane system at the CG level. 
For a molecular system, considering a \emph{d}-dimensional system with \emph{n} atoms in the canonical ensemble, the coordinates of the atoms are \emph{$\textbf{q} = {{q_1,q_2,...,q_{dn}}}$} $\in$ $\mathbb{R}^{dn}$ in the laboratory frame. 
The configurational distribution function can be defined as

\begin{equation}
p(\textbf{q})=\frac{1}{Z}e^{-\beta{V(\textbf{q})}},
\label{eq:eq1}
\end{equation}
where $V(q)$ is the interaction potential, $Z = \int{e^{-\beta{V(\textbf{q})}}}d\textbf{q}$ is the partition function, and $\beta=1/k_BT$.
The CG variables \textbf{$\xi$}(\textbf{q}) can be defined as a reduced set of atoms, i.e., {$\xi$}(\textbf{q}) = \{{$\xi_1$}(\textbf{q}), {$\xi_2$}(\textbf{q}), ..., {$\xi_k$}(\textbf{q})\},\emph{ k < dn}. 
The mass and momentum of CG variables are ${M_j}=\sum_{i=1}^{n}{m_{ji}}$ ($m_{ji}$ is the mass of the $i$ atom in the $j$ CG variable) and ${P_j}=\sum_{i=1}^{n}{P_{ji}}$ ($P_{ji}$ is the momentum of the $i$ atom in the $j$ CG variable), respectively. The configurational distribution of the CG system is the projection of the 
configurational distribution of the atomistic system onto the CG variables:

\begin{equation}
p(\textbf{$\xi$})=\frac{1}{Z}\int{e^{-\beta{V(\textbf{q})}}}\delta(\xi(\textbf{q})-\xi)d\textbf{q}
\label{eq:eq2}
\end{equation} 

The interaction potential energy and the force on the CG site can be defined by the above probability distribution function.
\begin{equation}
U(\textbf{$\xi$})=-\frac{1}{\beta}ln(p(\textbf{$\xi$}))
\label{eq:eq3}
\end{equation} 

\begin{equation}
F(\textbf{$\xi$})=-\nabla_{\xi}U(\textbf{$\xi$})
\label{eq:eq4}
\end{equation} 

\begin{figure*}
    \centering
    \includegraphics[width=6in]{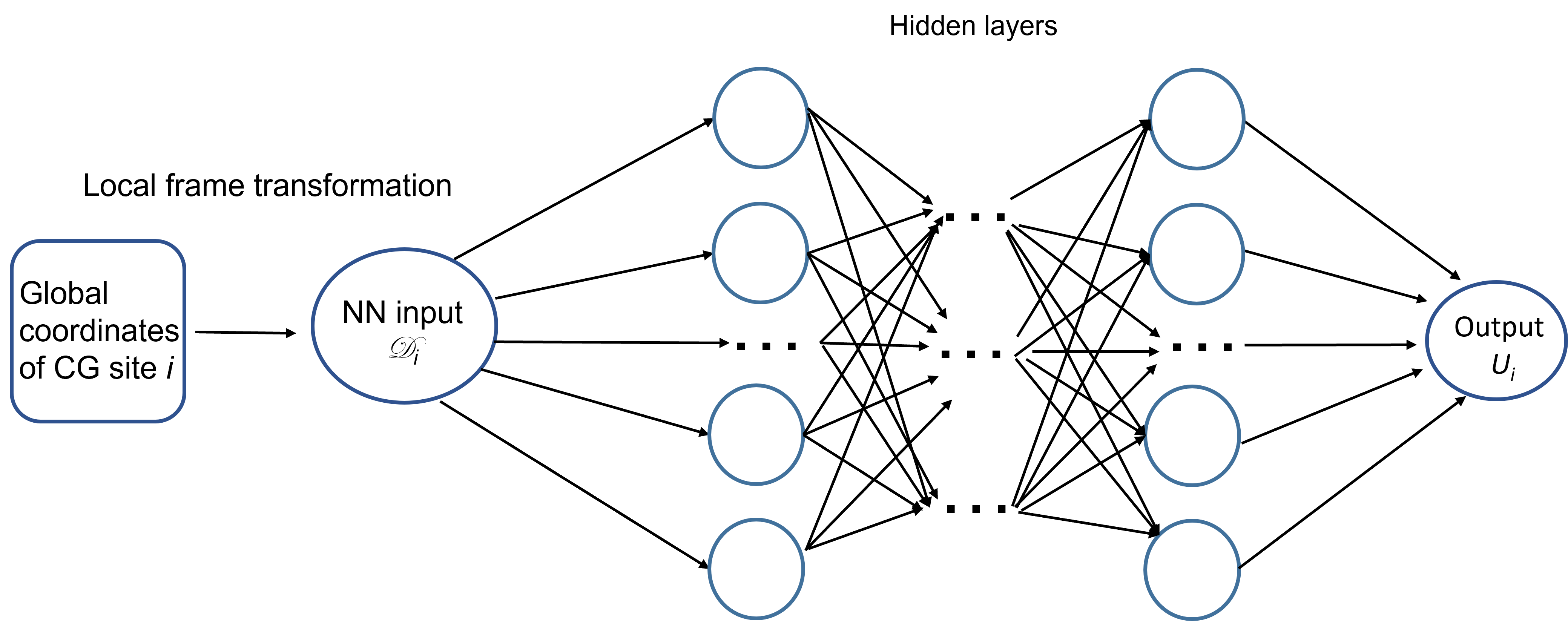}
    \caption{\textsf{The architecture of the first neural network.}}
    \label{fig:n1architecture}
\end{figure*}

Obviously, U(\textbf{$\xi$}) depends on the degree of coarse graining. In general, when we build a CG model 
from an atomistic simulation, three steps are needed: (1) determine the degree of coarse-graining and one or multiple quantities of interest (QoI); (2) 
select the analytical function(s) to represent the interaction potential between the CG sites; and (3) optimize the parameters in the analytical function(s) to reduce the difference of the values of the QoI(s) between the CG simulation results and the reference simulation results.
 Steps 2 and 3 can be merged for some methods that do not use analytical functions like iterative Boltzmann inversion (IBI) and inverse Monte Carlo (IMC). \cite{Reith2003,Lyubartsev2002} 
Depending on the QoIs, the CG methods can be classified as top-down method (using macroscopic properties as QoIs) and bottom-up method (using microscopic properties as QoIs).
For some properties, e.g., structural parameters, it may be difficult to provide the complete configurational distribution of the CG variables in the atomistic simulation. 
Usually only two- or three-body correlation functions like the radial distribution function and/or the angular correlation function are selected as descriptors.
 In the current work, a neural network representation $\tilde{U}$($\xi$,$\theta$) was adopted for the CG interaction potential, where
$\theta$ represents the neural network parameters. To simplify the model, one octane molecule is coarse-grained to one CG site. 
The position of the CG site is at the center of mass of the octane molecule. The derivation of a CG potential is achieved by training a ML model to predict the forces and/or energies of particles.
 Therefore, a force-matching scheme is typically used. The NN CG potential involves intrinsically many-body interactions. It should preserve the translational and rotational invariance as well as the permutational symmetry of the CG sites. 
The CG potential is based on a sum of the local contributions of the CG particles. We built the CG potential in two steps. First, the global coordinates $R$ of the CG particle system and its neighbors within a cut-off radius ${R}_{cut}$ were transformed into the descriptor with local frame transformation.\cite{Zhang2018,wang2020}  
The local coordinates
of the CG site \emph{I} were constructed by the positions of the CG site \emph{I} and its first- and second-nearest
neighbors. In the local coordinates, we can define the vector {$x_{ij}$, $y_{ij}$, $z_{ij}$} from CG site \emph{i} to CG site \emph{j}. The local descriptor matrix {$\tilde{R}_{ij}$} is given by:
\begin{equation}
\tilde{R}_{ij}=\{{s(R_{ij})},\frac{s(R_{ij})x_{ij}}{{{R}_{ij}}^2},\frac{s(R_{ij})y_{ij}}{{{R}_{ij}}^2},\frac{s(R_{ij})z_{ij}}{{{R}_{ij}}^2} \},
\label{eq:d1}
\end{equation} 
where $s$ is a smooth weighting function that decays to 0 at $R_{cut}$.\cite{ZhangNEURIPS2018}. This represents the local environment of the $i$ CG site to its $N_i$ neighbors within $R_{cut}$.  
Second, a local embedding network $G(s(R_{ij}))$ maps from a single value $s(R_{ij})$ to $N_1$ outputs. 
The network parameters of $G$ depend on the species of both the CG site $i$ and its neighbor $j$. The matrix form of $G(s(R_{ij}))$ is actually the 
local embedding matrix $\mathcal{G}^I\in\mathbb{R}^{{N_I}\times{N_1}}$,
\begin{equation}
(\mathcal{G}^{i})_{JK}=(G(s(R_{ij})))_K
\label{eq:d2}
\end{equation} 
Then, the encoded feature matrix  $\mathcal{D}^i\in\mathbb{R}^{{N_1}\times{N_2}}$ of the CG site $i$ can be defined as
\begin{equation}
\mathcal{D}^{i}=(\mathcal{G}^{i1})^{T}\tilde{R}^{i}(\tilde{R}^{i})^T\mathcal{G}^{i2}
\label{eq:d3}
\end{equation} 
The translational and rotational symmetries were preserved because the symmetric matrix $\tilde{R}^{i}(\tilde{R}^{i})^T$ is an over-complete array of invariants with respect to translation and rotation. The permutational symmetry was also preserved because the subnetworks associated with the same type of particles share the same parameters. The descriptors were given as input to a fully connected feed forward neural network to compute the potential contribution of the CG site. 
 The neural network structure is presented in Fig. \ref{fig:n1architecture}. 
 
We adopted a force-matching
scheme to train the NN potential of the CG model. The loss function is 
 \begin{equation}
\emph{L}=\frac{1}{\emph{CN}}\sum_{i=1}^{C}\sum_{j=1}^{N}\left|\textbf{F}_{i}(\tilde{R}^i)+\nabla_{\tilde{R}_i}\tilde{U}(\tilde{R}^i,\theta)\right|^2
\label{eq:loss1}
\end{equation} 
where C is the number of configurations in the training data. $i$ is the $i$ configuration.
$\textbf{F}_{i}$ is the instantaneous total force on the $i$-th CG site.
 
The $N_1 \times N_2$ components in the matrix $\mathcal{D}^{i}$ were reshaped into a vector and used as the input of NN1, yielding the target property. 
The DeePMD-kit library\cite {wang2018} was used as the interface to Tensorflow and the solver (LAMMPS)
for NN model training and calculation of the energy and force. In the NN, there are five hidden layers, and for each hidden layer the number of neurons decreases, given by 160, 80, 40, 20, and 10 from the first to the last hidden layer, respectively. 
 We used the hyperbolic tangent function as the activation function. The Adam stochastic gradient descent
(SGD) method was employed to optimize the loss function. The learning rate is adaptive. During the training, the learning rate decayed exponentially from the starting
learning rate 0.001 to the end learning rate 1 $\times$ $10^{-8}$ every 10,000 steps. As the
training cost increases with the numbers of neurons and layers, a neural network with fewer neurons and layers is
preferable.

\subsection{\label{sec:level24} The DPD framework}
Even if the CG interaction potential can match the reference energy and/or force, the equation of motion
at the CG level needs to include more terms so that it can reproduce the dynamical properties like the diffusion coefficient. The dynamics generated by molecular dynamics simulations at the atomic scale 
can be described at the CG level by a generalized Langevin equation (GLE).\cite{Kawasaki1973} 
\begin{equation}
\dot{\textbf{q}}_{i}=M_i^{-1}\textbf{P}_{i}
\label{eq:d01}
\end{equation}
\begin{equation}
\dot{\textbf{P}}_\emph{i}=\left\langle{\textbf{F}}_\emph{i}\right\rangle+\int_{0}^{t}{K(t-t')M_i^{-1}\textbf{P}_\emph{i}(t')dt'} +\widetilde{\textbf{F}}_\emph{i}
\label{eq:d02}
\end{equation}
where $\left\langle{\textbf{F}}_\emph{i}\right\rangle$ is the conservative force term determined by the potential of mean force. The third term is a random force. 
The second term involves a memory kernel $K(t-t')$, which is related to the random force by
$K(t) = \beta\left\langle{\widetilde{\textbf{F}}_{i}(t)\widetilde{\textbf{F}}_{i}(0)}\right\rangle$ to satisfy the second fluctuation-dissipation theorem. \cite{Li2017,Yoshimoto2017,Li2015}
The memory kernel can be derived from the microscopic Newtonian equations of motion by means of the
Mori-Zwanzig projection operator formalism. \cite{chorin2007problem} In previous applications, only the pairwise force was considered. However, because we used the NN potential to represent and calculate the conservative force in the DPD equation, the effect of many-body interaction is implicitly included. Although a couple methods were recently  proposed to determine the memory kernel with many-body interactions\cite{wang2020}, the mathematical formulation remains complicated, which makes the calculation slow.

The target molecule in the current system is a small molecule. Compared with complex fluids, its diffusion behavior is much simpler. 
Therefore, we optimized the CG dynamics based on the standard DPD approach. The DPD equation of force is given by
\begin{equation}
{\textbf{F}_i}={\textbf{F}_i^C}+{\textbf{F}_i^D}+{\textbf{F}_i^R}
\label{eq:d11}
\end{equation} 
where $F_I^C$ denotes the conservative force on CG site \emph{I}.
The nonconservative forces can be further partitioned into velocity-dependent dissipative friction and
stochastic terms, which satisfy the fluctuation-dissipation theorem. 
The dissipative force and random force can be split into pair forces between beads as
\begin{equation}
\textbf{F}_i^D=\sum_{i\neq{j}}{\textbf{F}_{ij}^D}
\label{eq:d12}
\end{equation}
\begin{equation}
\textbf{F}_i^R=\sum_{i\neq{j}}{\textbf{F}_{ij}^R}
\label{eq:d21}
\end{equation}
where the dissipative force is
\begin{equation}
\textbf{F}_{ij}^D=-{\gamma^\parallel}\emph(w^D)(r_{ij})(\hat{r}_{ij}\cdot\vec{v_{ij}})\hat{r}_{ij}
\label{eq:d31}
\end{equation} 
and the random force is given by
\begin{equation}
\textbf{F}_{ij}^R={\sigma^\parallel}\emph(w^R)(r_{ij}){\Theta/\sqrt\Delta{t}}\hat{r}_{ij}
\label{eq:d4}
\end{equation}
where $\gamma^\parallel$ and $\sigma^\parallel$ are the friction coefficient and the noise strength of the generalized thermostat along the interatomic axis, respectively. $\Theta$ is a Gaussian random number with zero mean value and unit variance. 

$(w^D)(r_{ij})$ and $(w^R)(r_{ij})$ are $r$-dependent weight functions.

Equations \ref{eq:d31} and \ref{eq:d4} can be generalized as
\begin{equation}
\textbf{F}_{ij}^D=-{\gamma}\emph(w^D)(r_{ij}){\textbf{P}}_{ij}({r}_{ij}){v_{ij}}
\label{eq:d5}
\end{equation}
 and 
\begin{equation}
\textbf{F}_{ij}^R={\sigma}\emph(w^R)(r_{ij}){\textbf{P}}_{ij}({r}_{ij}){\Theta/\sqrt\Delta{t}},
\label{eq:d6}
\end{equation}
where ${P}_{ij}({r}_{ij})$ is a projection operator.
If we choose the projector along the interatomic axis between CG sites \emph{i} and \emph{j}, it is the standard DPD.
\begin{equation}
{\textbf{P}}_{ij}({r_{ij}})=\hat{r}_{ij}\otimes\hat{r}_{ij}
\label{eq:d7}
\end{equation}
One can project on the plane perpendicular to the interatomic axis. The space defined by the following projector is perpendicular to the case of the standard DPD.
\begin{equation}
{\textbf{P}}_{ij}({r_{ij}})={I}-\hat{r}_{ij}\otimes\hat{r}_{ij}
\label{eq:d8}
\end{equation}
The two friction constants ${\gamma^\parallel}$ and ${\gamma^\perp}$ can be adjusted in the transverse DPD thermostat. For the transverse DPD thermostat, Galilei invariance remains valid by construction. To reproduce the dynamical property of the fluid at the CG scale, we can tune the appropriate friction coefficients in the DPD equation while also reproducing the key structural properties of the fluid.

In this work, we also tested the transverse DPD thermostat because it has been reported that the viscosity is very sensitive to the damping perpendicular to the interatomic axis in DPD simulation. \cite{Junghans2008}

\subsection{\label{sec:level25}The second neural network}
In the previous sections, we built a DPD model of octane fluid that can reproduce the liquid structure and dynamic property by integrating NN potential.
The model was further used in the simulations of constrained fluid flow in a channel. The effect of the boundary shape was investigated in the constraint fluid flow simulation with walls. To simplify the model, the walls consist of particles that have the same interaction as the particle in the fluid, i.e., the "very hydrophilic" surface. These wall particles are fixed during the simulation,\cite{yong2013} and the packing density of particles is larger than in the fluid, like in a body-centered cubic (BCC) lattice. 
The boundary of the the channel in the simulations was mimicked as a sine function. The shape can be tuned by adjusting the coefficients in the sine function as well as the curvature. 
Two methods are widely used to observe and analyze fluid flows. One method involves observing the trajectories of specific fluid parcels, which yields what is commonly termed a Lagrangian representation.
The other method includes observing the fluid velocity at fixed positions, which yields an Eulerian representation. 
Lagrangian methods are often the most efficient way to sample a fluid flow, and the physical conservation laws are inherently Lagrangian because they apply to moving fluid volumes rather than to the fluid
that happens to be present at some fixed point in space.
Nevertheless, the Lagrangian equations of motion applied to a three-dimensional continuum are quite difficult in most applications, and thus almost all of the theory (forward calculation) in fluid mechanics is developed within the Eulerian system.
Both methods are commonly used in the analysis of fluid mechanics. Actually, the transformation of the Eulerian velocity field from a Lagrangian velocity field is necessary in most fluid dynamics simulations because Lagrangian methods are natural for many observational techniques and for stating the fundamental conservation theorems. Almost all of the theory in fluid mechanics has been developed in the Eulerian system. For this reason, we have to consider both coordinate systems and the transformation from one to the other. An important and rapidly developing observational technique involves generating the Eulerian velocity field from Lagrangian data through the analysis procedure of interpolating or mapping irregularly sampled Lagrangian data to a spatial grid. This usually needs lots of samples to get an accurate result. For the particle-based simulation, it is not cheap to perform many simulations for velocity data generation. The three-dimensional Lagrangian equations in fluid mechanics are usually very complex, and the Eulerian equations are always used instead.

\begin{figure*}
    \centering
    \includegraphics[width=6in]{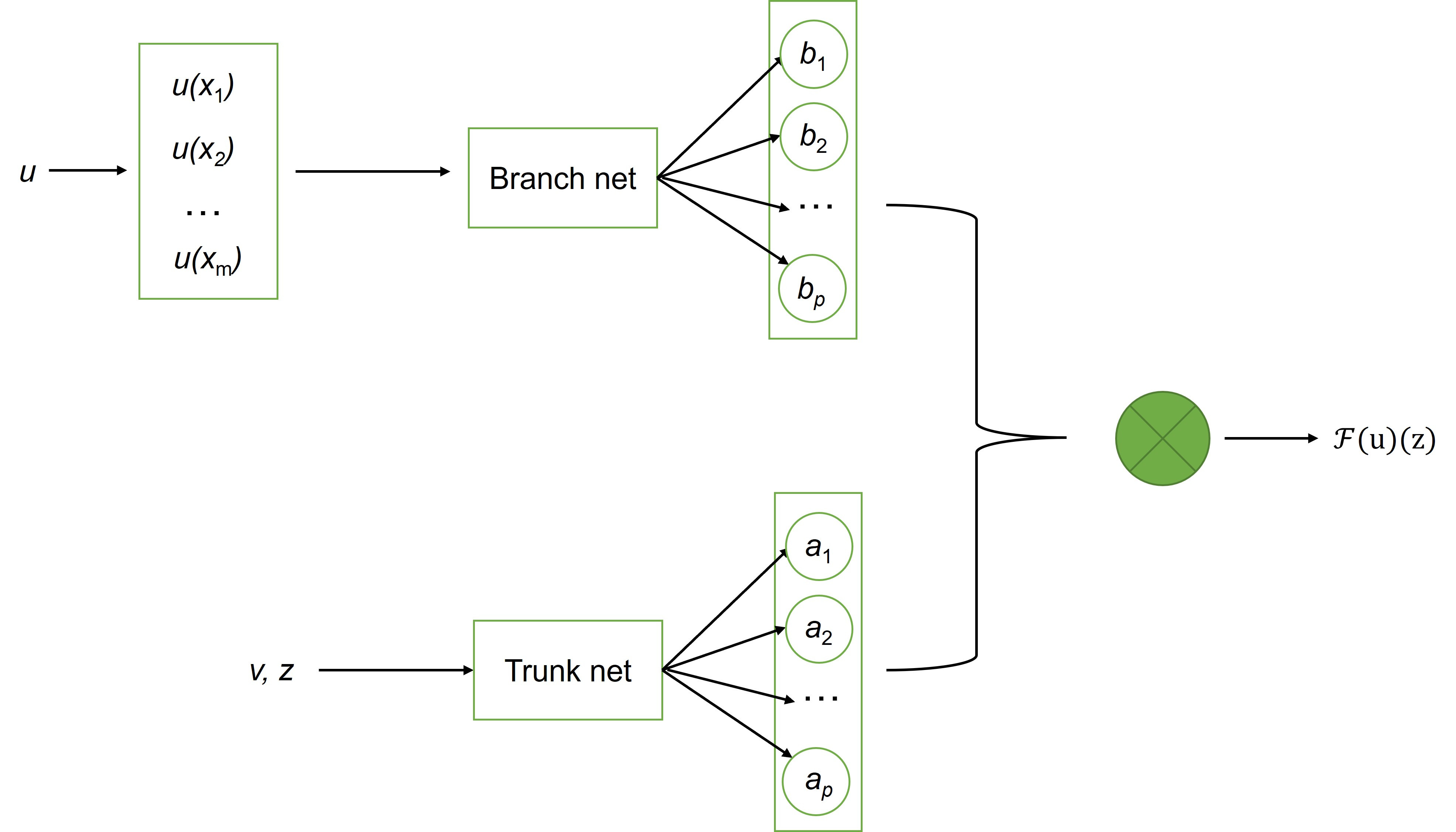}
    \caption{\textsf{The architecture of the second neural network.}}
    \label{fig:n2architecture}
\end{figure*}

Deep learning algorithms have been applied to address challenging fluid problems. Here we used a deep NN (DNN) model to predict the velocity profile of the constrained flow. 
Recently, learning operators mapping between infinite-dimensional function spaces by DNNs has become a new paradigm. This type of DNN is called deep neural operator. Here, we used the architecture of a deep neural operator called deep operator networks (DeepONet).\cite{Lu2021,Lu2022,howard2022multifidelity}
A DeepONet consists of a network (called “branch net”) for
encoding the input function space and another network (called
 “trunk net”) for encoding the domain of the output functions. The DeepONet is based on the universal function approximation
theorem that a NN with a single hidden layer can accurately approximate any nonlinear continuous functional. When considering a dynamical system governed by ordinary or partial differential equations, its state-space modeling can also be comprehended through the lens of the operator. Consequently, operator learning holds substantial potential for constructing state and space functions in a data-driven manner.
Additionally, it has been proved that the DeepONet architecture is extremely efficient. \cite{Cai2021} In this work, we used the unstacked DeepONet. The architecture is shown in Fig. \ref{fig:n2architecture}. The branch and the trunk nets were trained simultaneously.
With the NN model (called NN2 in this work), we are able to predict the velocity profile of the constrained fluid.
The input to the branch network is the boundary $u$ in our simulation system, discretized at points ${x_i}$ ($i = 1, 2, ..., m$). The output is $\left[ b_1, b_2, ..., b_p\right]^T$ $\in$ $\mathbb{R}^{p}$. 
The input to the trunk net is the position $z$ and the velocities $v \in$ $\mathbb{R}^{kd}$, and the trunk output is $\left[ a_1, a_2, ..., a_p\right]^T$ $\in$ $\mathbb{R}^{p}$. The intermediate outputs are combined using a dot product, resulting in a DeepONet solution operator prediction.
\begin{equation}
\mathcal{F}\left(u\right)\left(z\right)=\sum_{k=1}^{p}{b_k}{a_k}
\label{eq:dd}
\end{equation} 
In terms of the loss function, we utilized the mean squared error (MSE), which measures the square of the $L_2$ norm between model predictions and the training velocity data along the flow. As part of the backpropagation process, the gradient of the loss function with respect to the weights in both networks is computed. This gradient is utilized by the Adam optimizer to modify the weights and minimize the loss value. Through a sufficient number of feedforward and backpropagation iterations, DeepONet learns the solution operator.

\begin{figure*}
    \centering
    \includegraphics[width=6in]{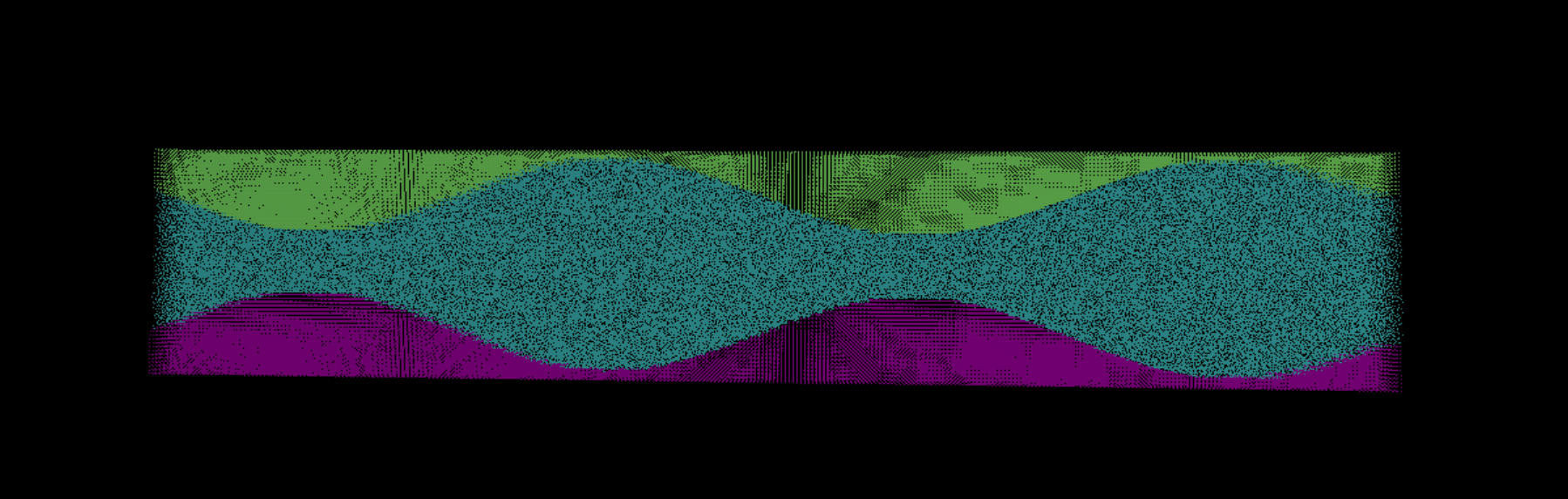}
    \caption{\textsf{The constrained fluid simulation system.}}
    \label{fig:system}
\end{figure*}

The training data were obtained via simulations with the above DPD model. In the DPD simulations, we performed nonequilibrium simulations to generate a Poiseuille flow by adding acceleration to the fluid particles in a channel. 
A snapshot of the simulation system is shown in Fig. \ref{fig:system}. The flow direction is $x$, and the flow is from left to right. 
The $x$ and $y$ directions of the system are periodical, while the $z$ direction is not periodic.
Note that here the different colors of the upper and lower boundaries in Fig.\ref{fig:system} indicate that their properties could be different (asymmetric boundary conditions). In our current simulations, they are just made of the same particles.
The shape of the boundary is controlled by a shifted sine function $\emph{y}$ = $a\sin\left(kx\right)+b$. In our simulations, to investigate the effect of the boundary shape on the velocity profile, $a$ and $b$ were fixed and $k$ was adjusted from 0.0125 to 0.2. We selected five $k$ values, i.e., 0.0125, 0.025, 0.05, 0.1, and 0.2.
For each $k$ value, 20,000 data frames were generated by DPD simulations. These data were fed to the DeepONet for velocity profile prediction. The Tensorflow\cite{abadi2016tensorflow} and DeepXDE\cite{Lu2021xde} libraries were used to build and train the neural networks.     

\section{\label{sec:level3}Result and discussion}

\subsection{\label{sec:level26}The first neural network: thermodynamic and structural property}
\begin{figure*}
\centering
\includegraphics[width=4in]{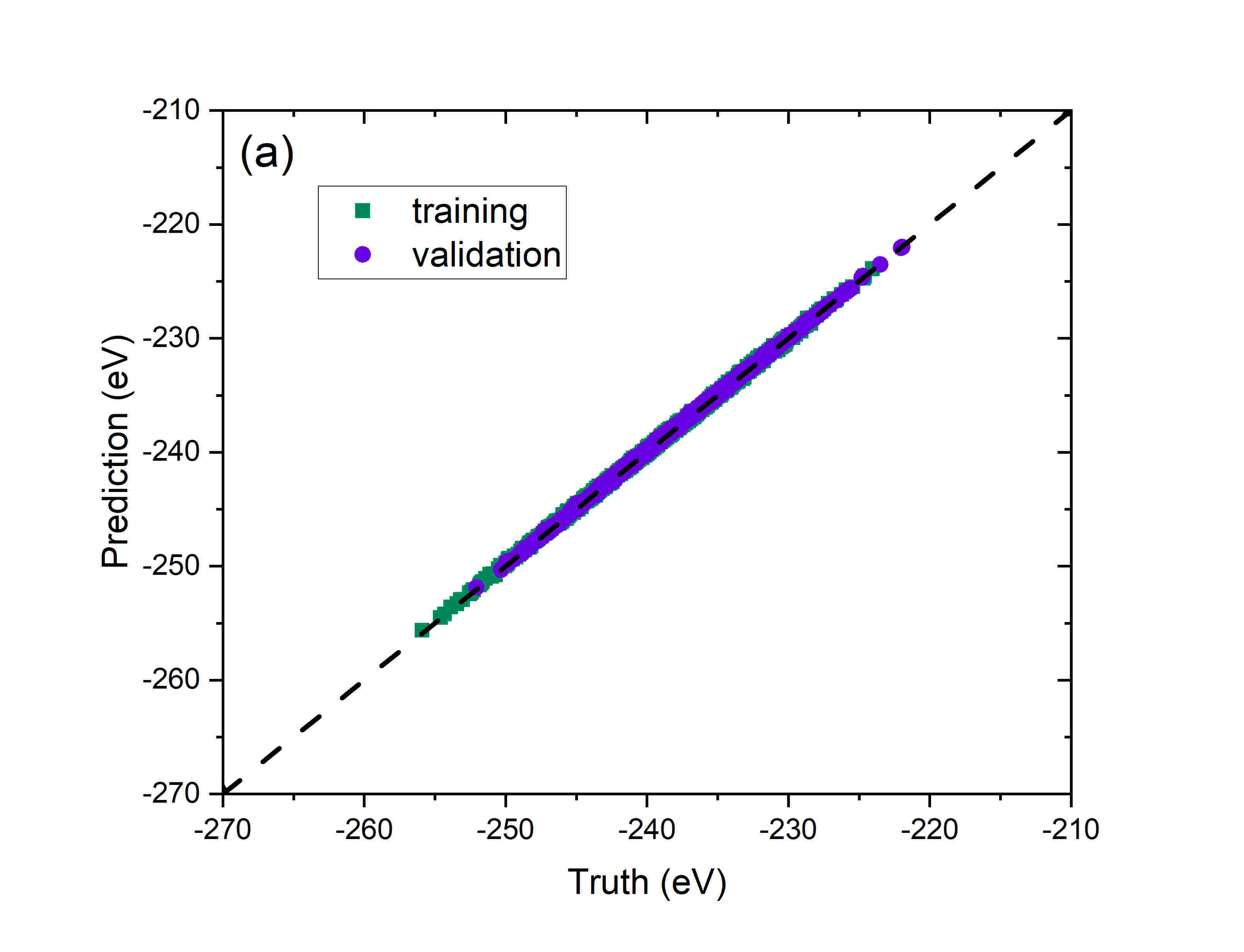}
\includegraphics[width=4in]{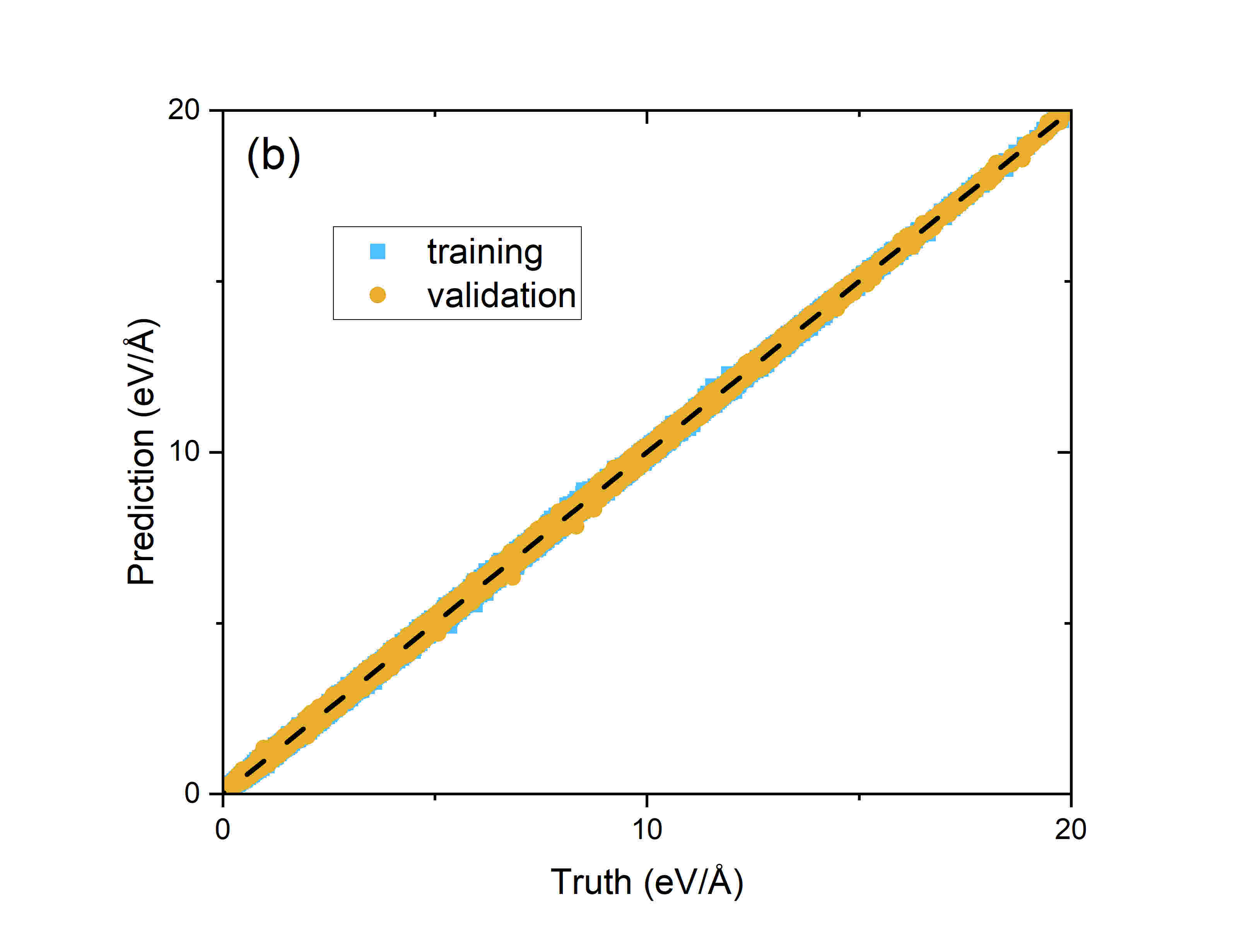}
\caption{\textsf{(a) The total energies of the simulation system in the training and validation datasets computed with the NN potential and reference energies. 
(b) The absolute value of the net force on the particle in the training and validation datasets computed with the NN potential and reference forces. The dash line represents a perfect fit.}}
\label{fig:energyandforce}
\end{figure*}

\begin{figure*}
\centering
\includegraphics[width=4in]{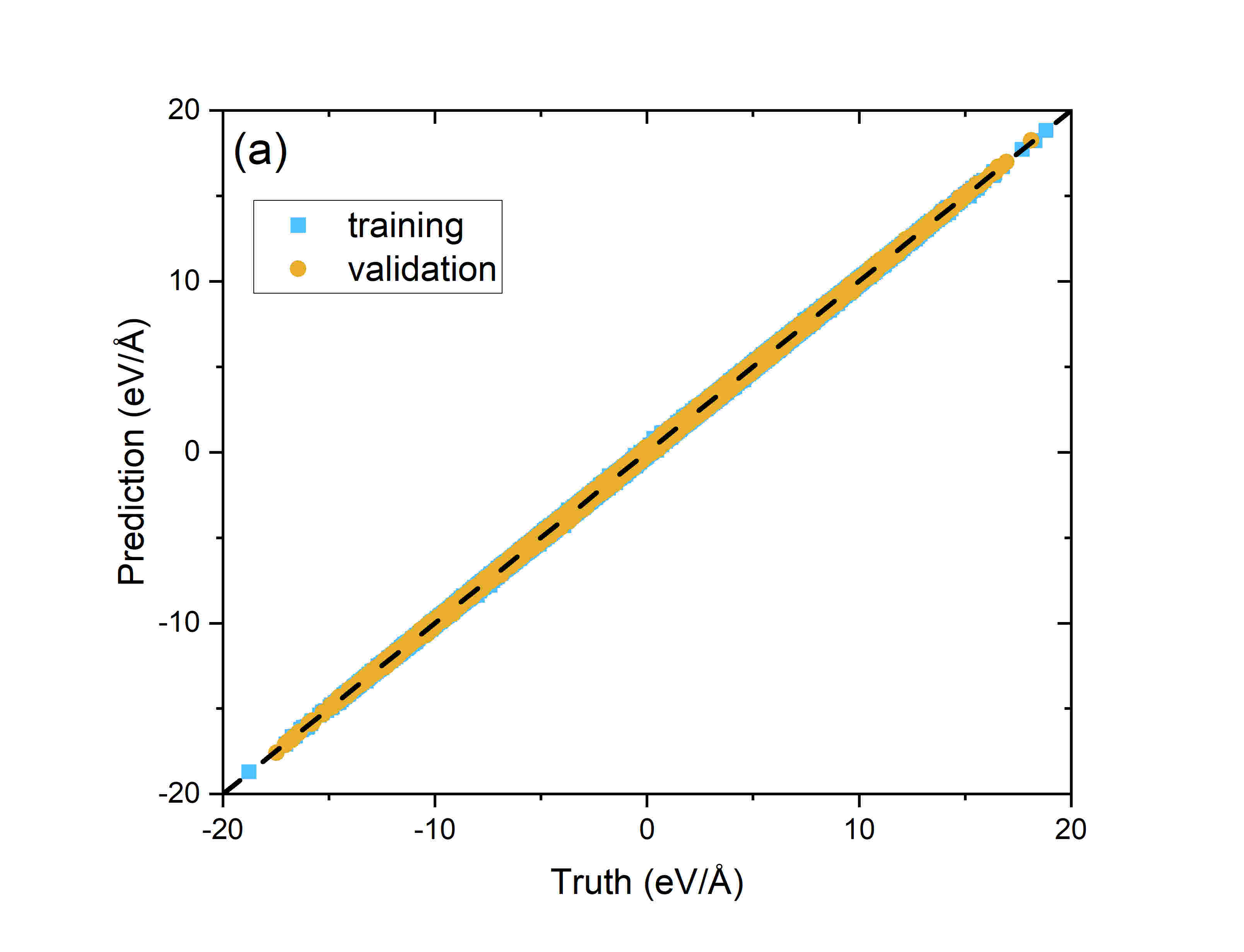}
\includegraphics[width=4in]{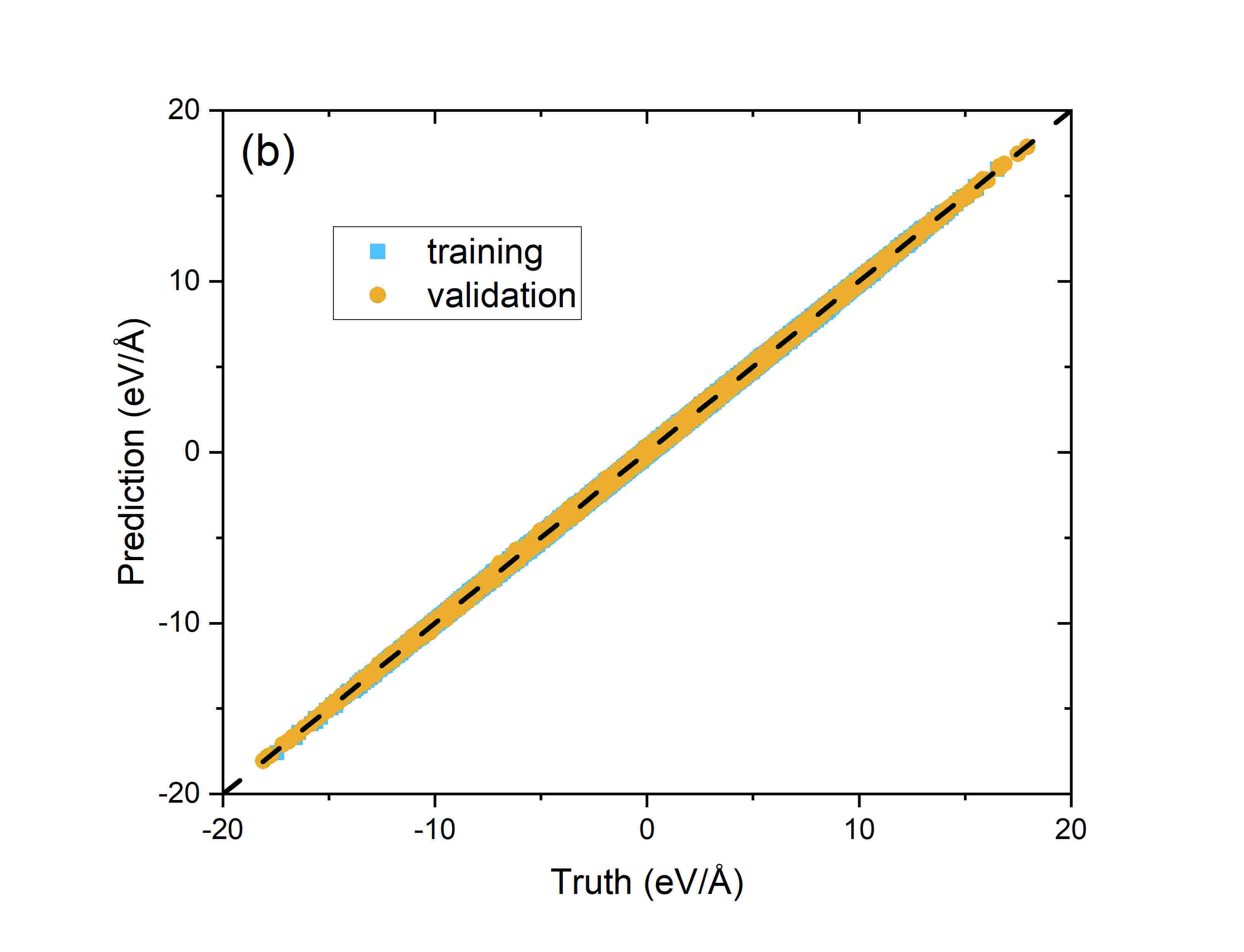}
\includegraphics[width=4in]{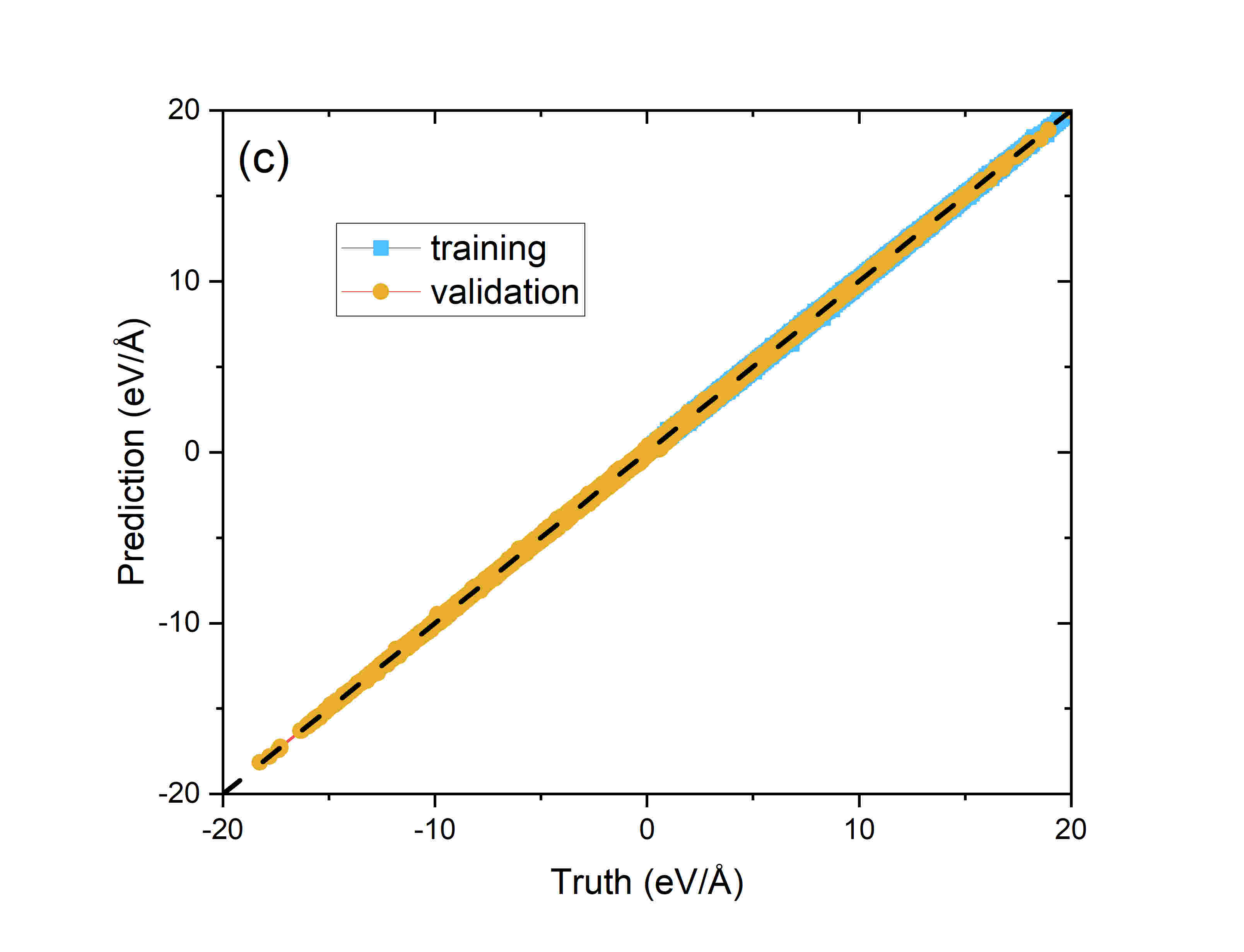}
\caption{\textsf{The force components on the particle in the training and validation datasets computed with the NN potential and references. (a) x direction. (b) y direction. (c) z direction. The dash line represents a perfect fit}}
\label{fig:forcecomponent}
\end{figure*}

\begin{figure*}
\centering
\includegraphics[width=4in]{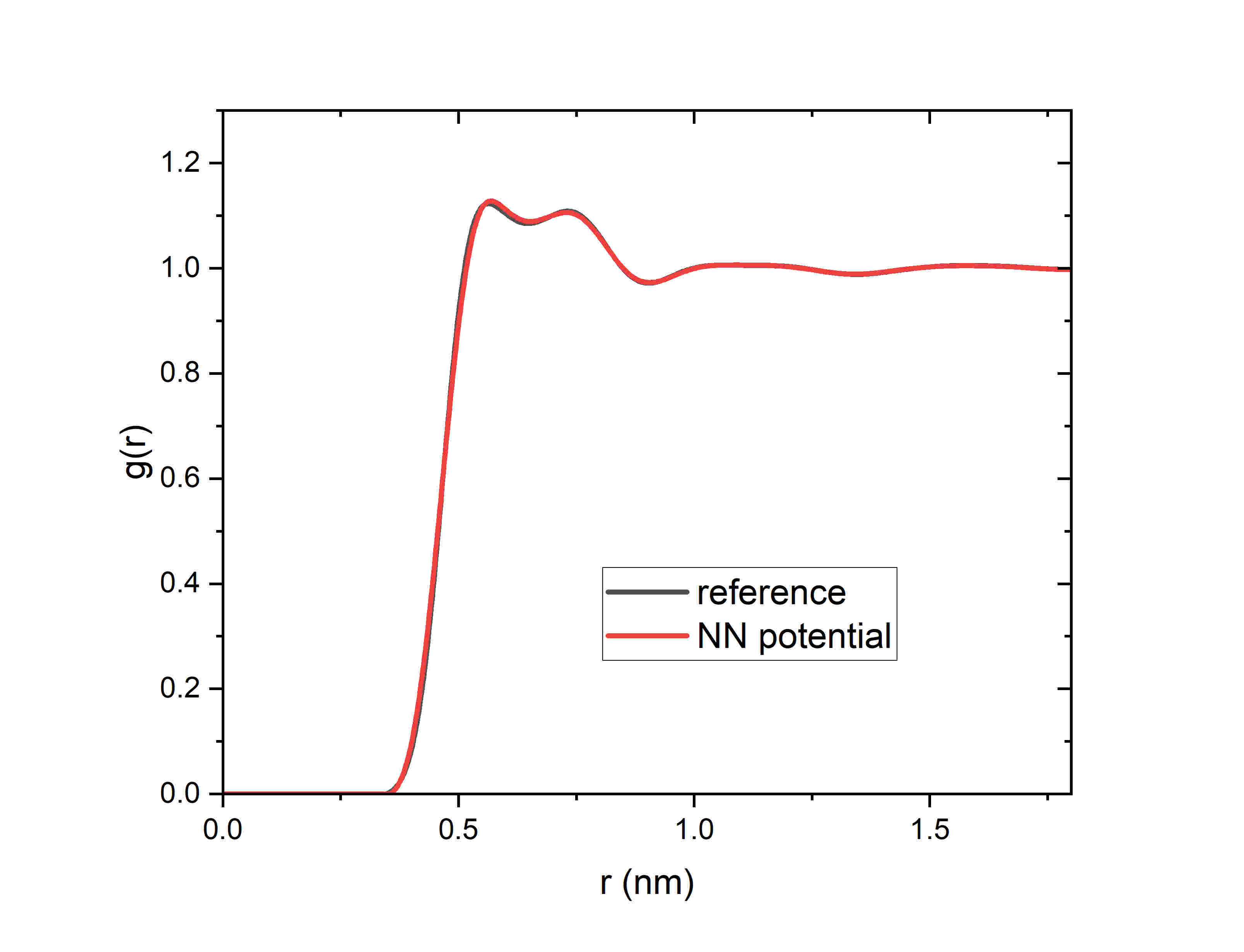}
\caption{\textsf{The radial distribution functions computed by the NN potential and reference. }}
\label{fig:rdf}
\end{figure*}

We trained a CG model of octane with ML potential, which is a type of liquid alkane at room temperature and the main component of gasoline. In the CG model, each octane is coarse-grained as a single CG site. The cutoff radius $R_c$ is set to 8.5 for the first neural network in the construction of the CG potential, which is about the position of the first valley in the radial distribution function (RDF). 
The CG simulation was numerically solved with the ML potential, and the energies, force, and structural property were computed. 
Figure \ref{fig:energyandforce}a demonstrates good correlation between the ground truth energies and the prediction energies over a 30 eV wide energy range for both potentials.
The root-mean-square errors (RMSEs) of the total energy for the training set and the validation set are 0.179 and 0.139 eV, respectively. The energies of the training set and validation set are close, indicating that the NN model is not overfitting.
The RMSEs of the total energy per particle are only 2.23 $\times$ $10^{-4}$ eV and 1.74 $\times$ $10^{-4}$ eV for the training set and validation set, respectively.
The correlation between the predicted forces and the reference is presented in Figure \ref{fig:energyandforce}b. 
The RMSEs of force between the prediction and ground truth are 0.042 eV/\si{\angstrom} and 0.044 eV/\si{\angstrom}, respectively. That demonstrates the accuracy of our NN potential.
The predicted force and ground truth results are also close, which further validated that our NN model is not overfitting.
The model prediction accuracy on the components of the force was also checked. The results of the three force components are shown in Figure \ref{fig:forcecomponent}.
The RMSEs at the three directions for the training set are 0.0421 eV/\si{\angstrom},
0.0420 eV/\si{\angstrom} and 0.0420 eV/\si{\angstrom}, respectively. These values are very close, indicating that the deviation is not due to a certain component of the force vector.
The RMSE for the validation set is similar. This also implies the homogeneity of our system because it is a liquid system and there is no artificial effect such as phase separation during simulation.
The structural property prediction result of the CG simulation using the NN potential was compared with the MD simulation. Fig.\ref{fig:rdf} shows the RDFs of the CG simulation with the NN potential and the reference. 
We can see that the CG model with machine learning potential can effectively reproduce the RDF in an atomic MD simulation. There are two peaks in the RDF plot because the octane molecule has two primary conformations in the atomic MD simulation.
As the structural property of the CG system is determined by the CG potential, the result demonstrates the employed NN method for building the CG potential.
This is difficult to be reproduced with only the two-body interaction potential.
Because the NN potential method naturally accounts for many-body interactions during construction of the CG potential, good performance can be expected, especially for cases of some molecular systems where many-body interactions are more significant than the pairwise interaction.

\subsection{\label{sec:level27}The first neural network: dynamical property}
\begin{figure*}
\centering
\includegraphics[width=3.5in]{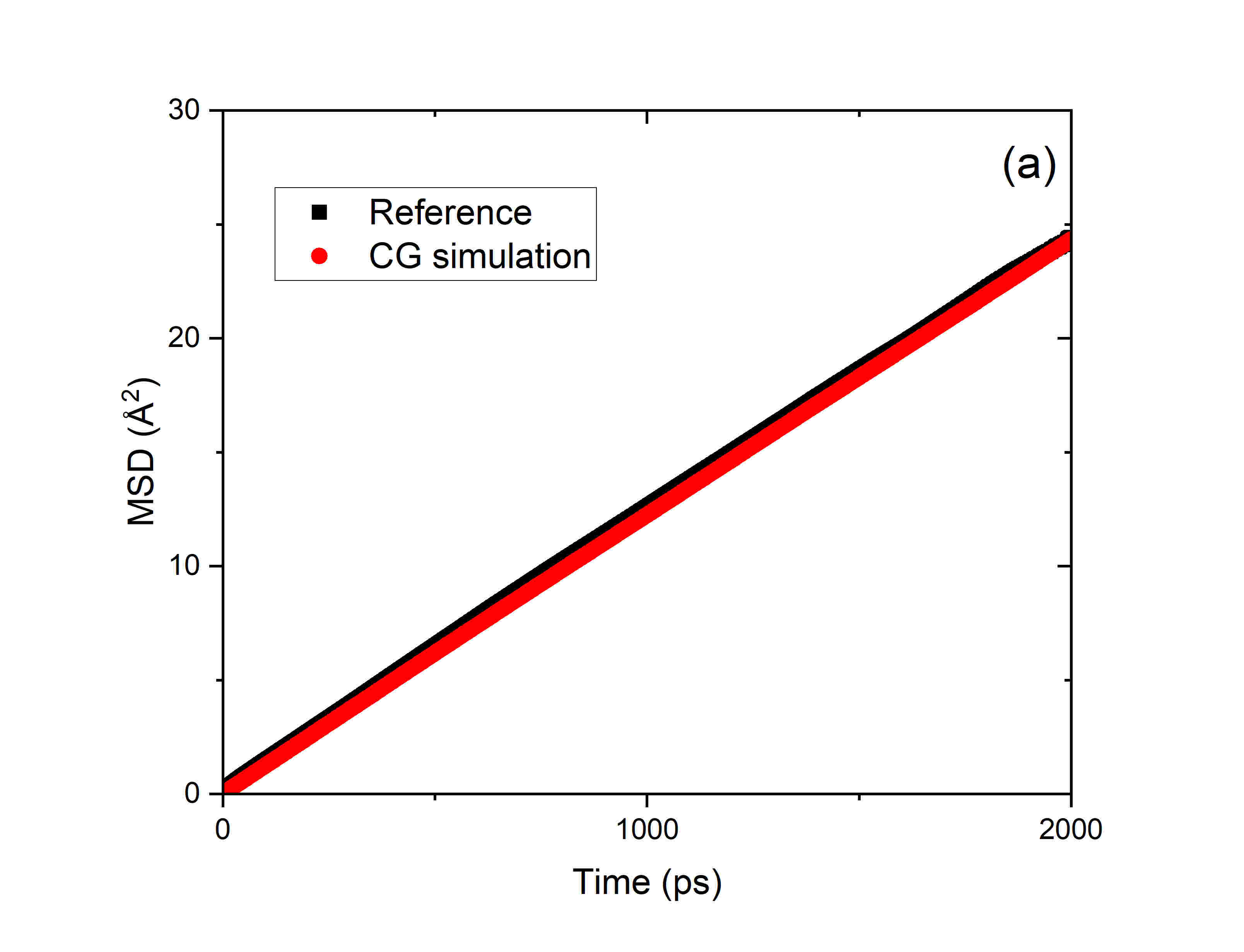}
\includegraphics[width=3.5in]{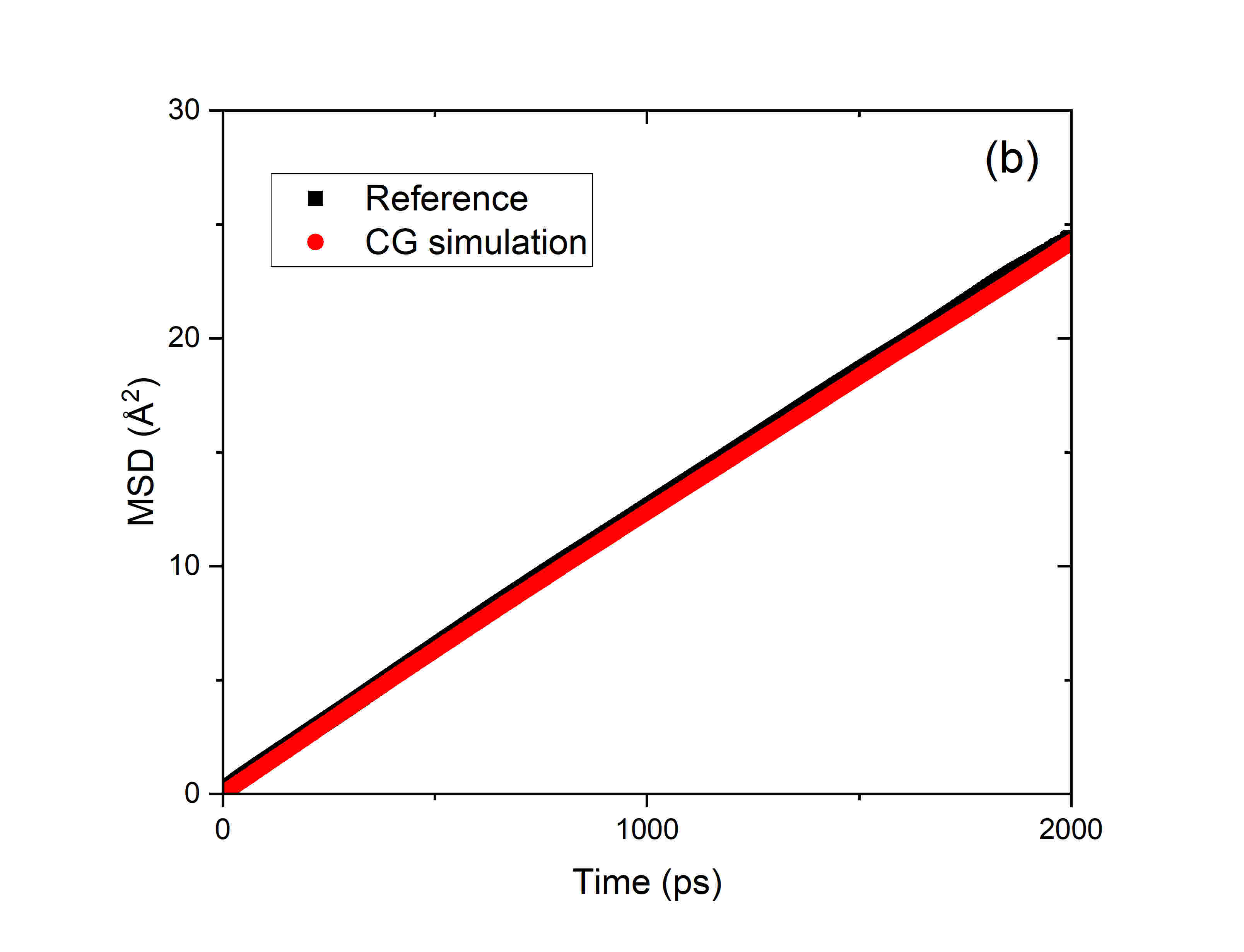}
\caption{\textsf{The mean square distance computed by the CG simulation and reference. (a) Standard DPD thermostat. (b) Transverse DPD thermostat.}}
\label{fig:msd}
\end{figure*}

Due to the loss of degrees of freedom in the CG model, the dynamics of the CG model are usually faster than the fine-grained (FG) model at atomic scale.
In a CG model with NN potential, although the NN potential can effectively reproduce the internal energy at the atomic scale, the loss of entropy such as conformation entropy is still a problem.
Therefore, the dynamics of the CG model with NN potential remain faster than the FG model.
To reproduce the dynamics of particles at CG level, the dynamics of the CG system can be further tuned in the framework of GLE or DPD when the conservative force part is obtained from the FG simulations.
It has been shown that for a system close to equilibrium,
the fluctuation–dissipation theorem can be recovered if considering an additive correction while
accounting for the local mean velocity of the particle. \cite{Jung2021,Speck2009} Because the properties
of the system at conditions far from equilibrium state are not our target,
we will not extensively cover the fluctuation–dissipation theorem of non-equilibrium systems.
It has been demonstrated that for a simple Lennard-Jones particle fluid, the DPD thermostat allows both diffusion coefficient and shear viscosity in CG
simulations to match the results in atomistic simulations even if the friction coefficient is fixed to the same value in both equilibrium and nonequilibrium CG simulations. \cite{Fu2013}
In this work, we integrated the CG model with NN potential into a DPD framework to further tune the dynamical property of the system.
We tried both the standard and transverse DPD thermostats. For the standard DPD thermostat, we tuned the friction coefficient in the DPD simulation to match the diffusion coefficient in the MD simulation. 
For the transverse DPD thermostat, the parallel friction coefficient and the perpendicular friction coefficient were tuned simultaneously.
Note that it is reported that the dynamical properties of fluid are more sensitive to the perpendicular friction coefficient.
However, to reduce the parameter space,
these two friction coefficients were kept consistent during optimization of the friction coefficient. This is because the solution is not unique for only matching the diffusion coefficient. More constraints are needed to obtain a unique solution for the friction coefficient in the transverse DPD thermostat. 
Figure \ref{fig:msd} shows the mean square displacement (MSD)
curves predicted by the CG
simulation and the reference. We can see good MSD agreement between the  simulation with NN potential and the MD simulation results for both standard and transverse DPD thermostats, 
which demonstrate our framework by reproducing the dynamic property of the reference system. Note that our system is large enough to avoid the hydrodynamic finite-size effect. \cite{Simonnin2017}
The viscosity is very sensitive to the perpendicular friction coefficient.
To obtain the same diffusion coefficient in the CG simulation, we found that the friction coefficient in the standard DPD thermostat must be 19 times greater than in the transverse DPD thermostat.
Because a too large friction coefficient may make some properties of the fluid unrealistic in a simulation, we used the transverse DPD thermostat in the following work.

\subsection{\label{sec:level28}The second neural network: hyperparameter optimization for velocity profile prediction}
\begin{figure*}
    \centering
    \includegraphics[width=4in]{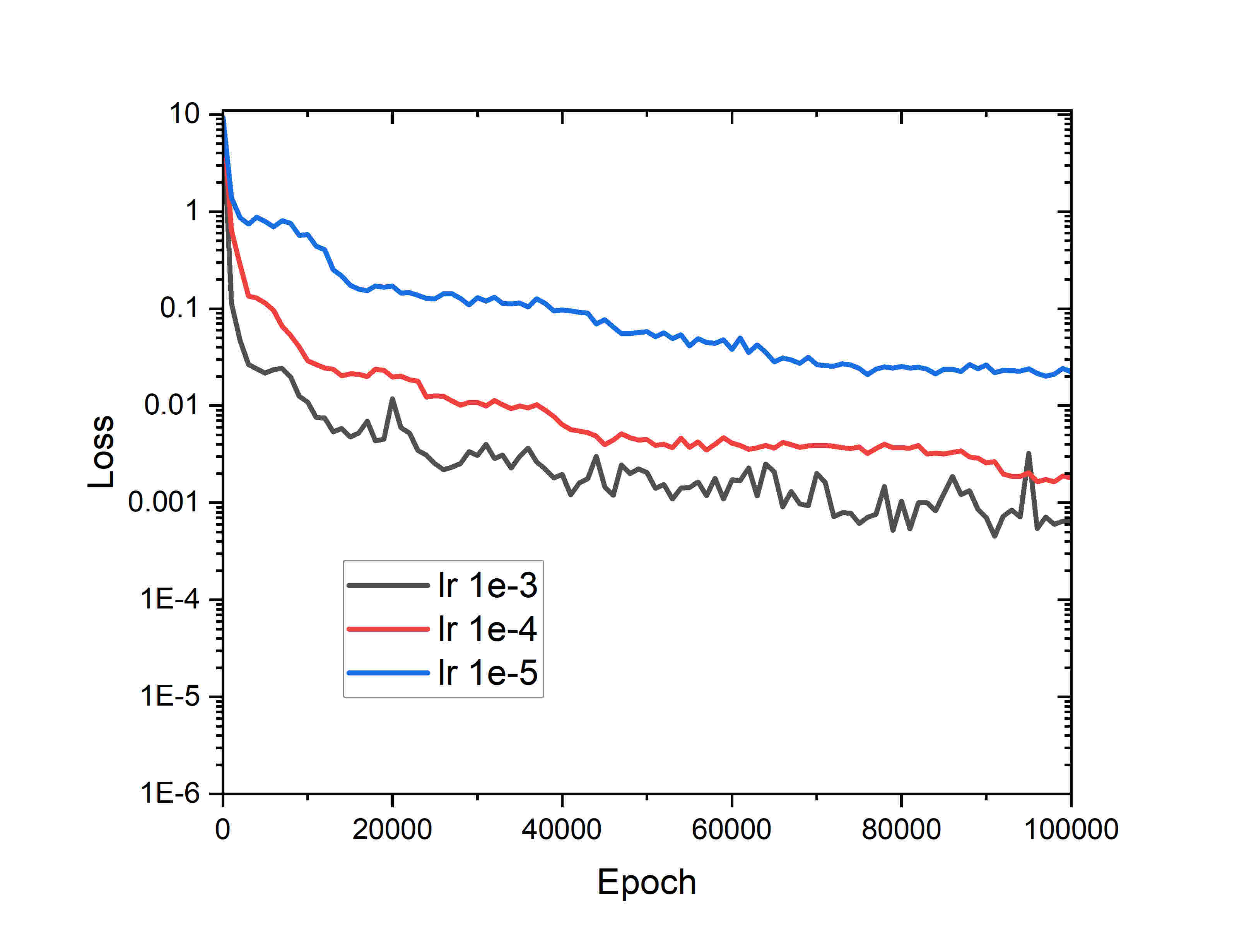}
    \caption{\textsf{Loss function with various learning rates as a function of epochs for NN2.}}
    \label{fig:hyperpara_lr}
\end{figure*}

\begin{figure*}
    \centering
    \includegraphics[width=4in]{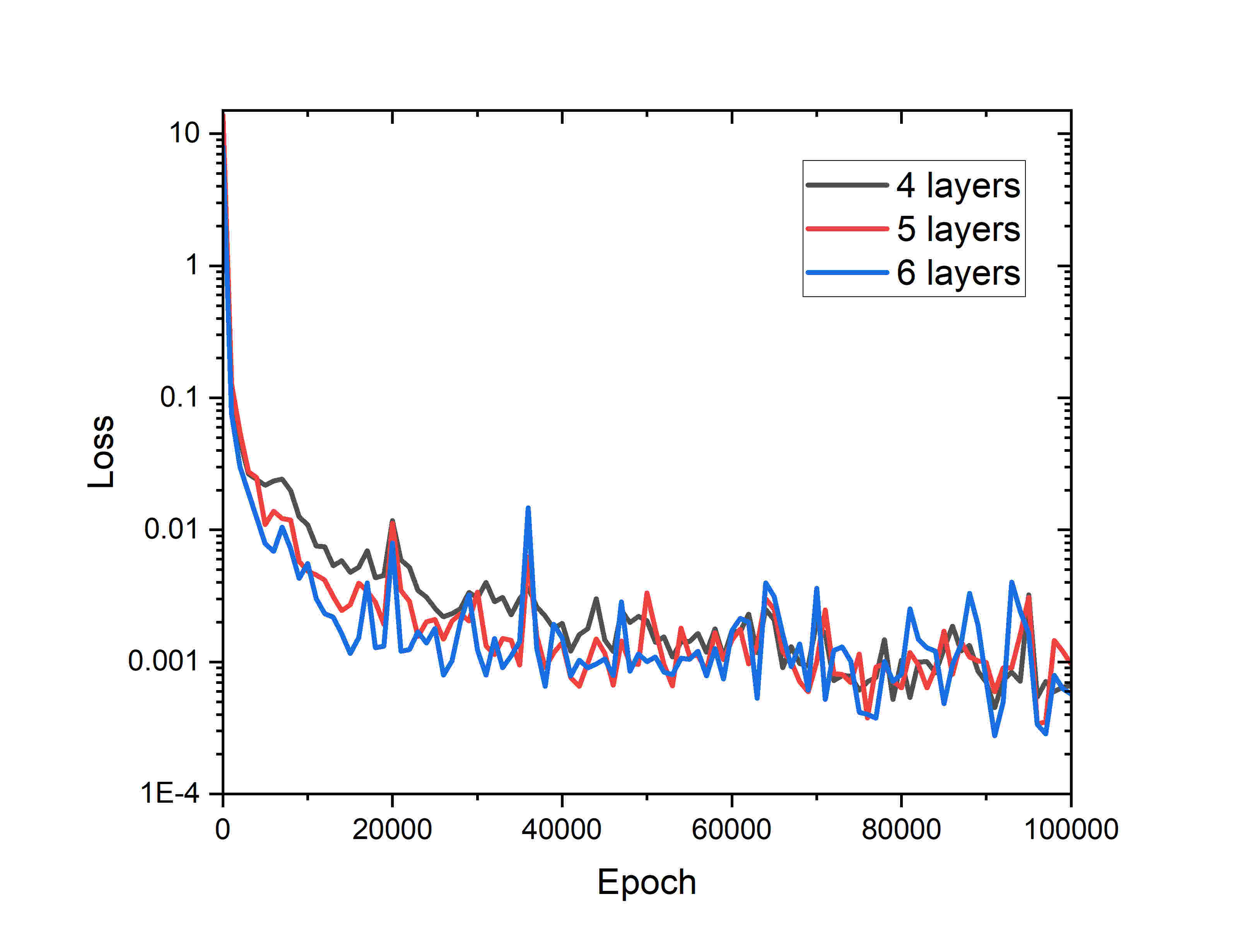}
    \caption{\textsf{Loss function with various NN layers as a function of epochs for NN2. }}
    \label{fig:hyperpara_NNlayer}
\end{figure*}

\begin{figure*}
    \centering
    \includegraphics[width=4in]{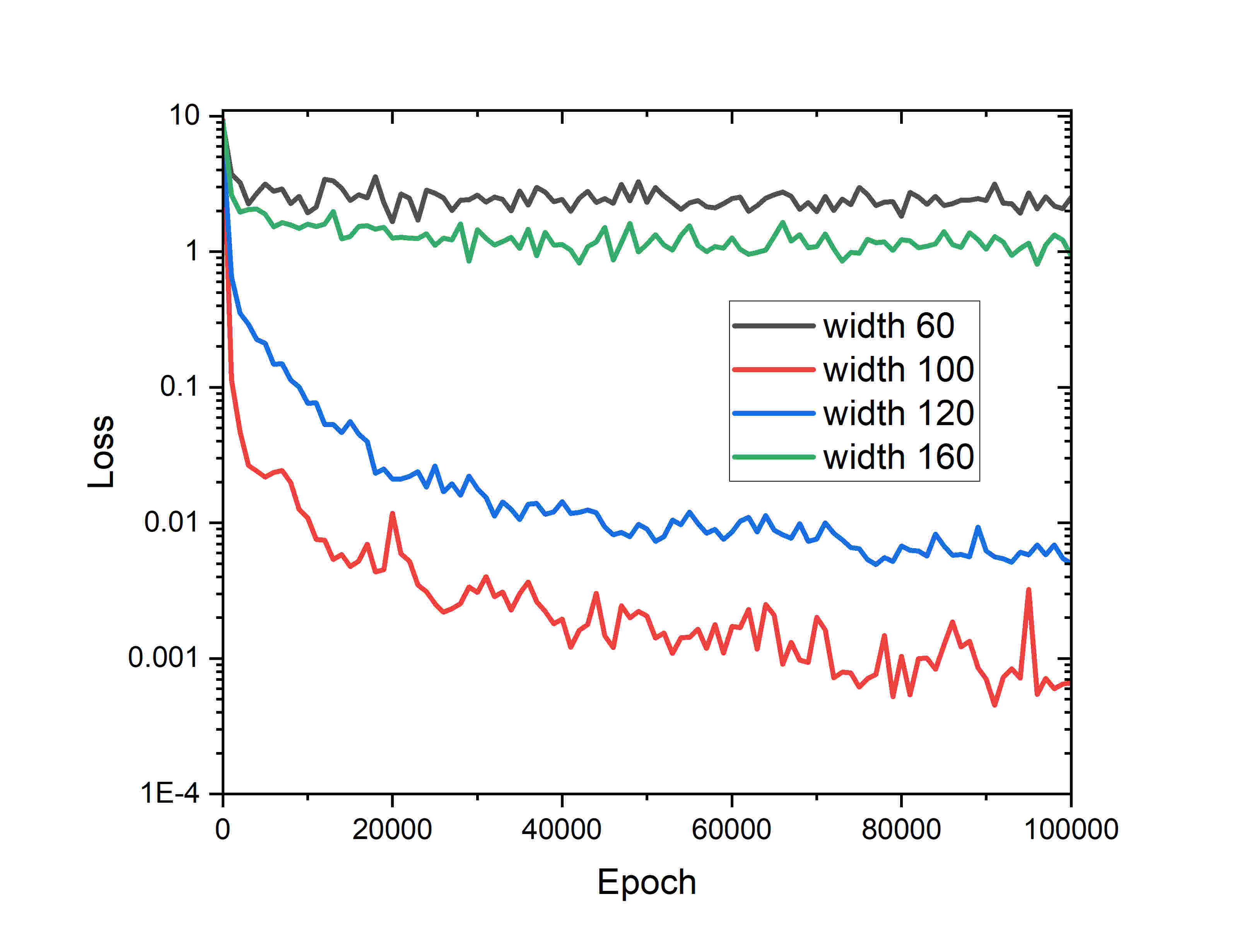}
    \caption{\textsf{Loss function with various NN widths as a function of epochs for NN2. }}
    \label{fig:hyperpara_NNwidth}
\end{figure*}

\begin{figure*}
    \centering
    \includegraphics[width=4in]{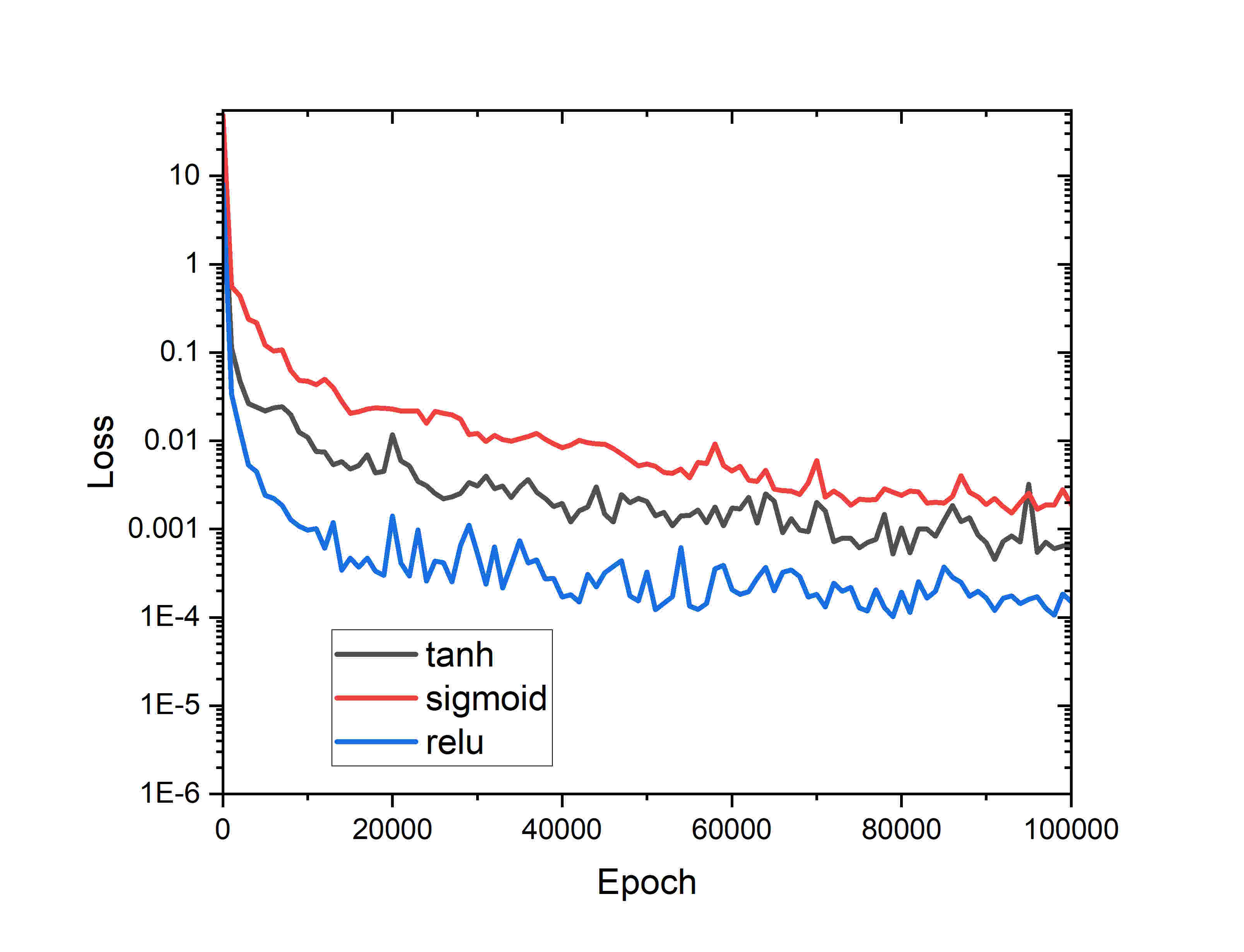}
    \caption{\textsf{Loss function with various activation functions as a function of epochs for NN2. }}
    \label{fig:hyperpara_af}
\end{figure*}

\begin{figure*}
    \centering
    \includegraphics[width=4in]{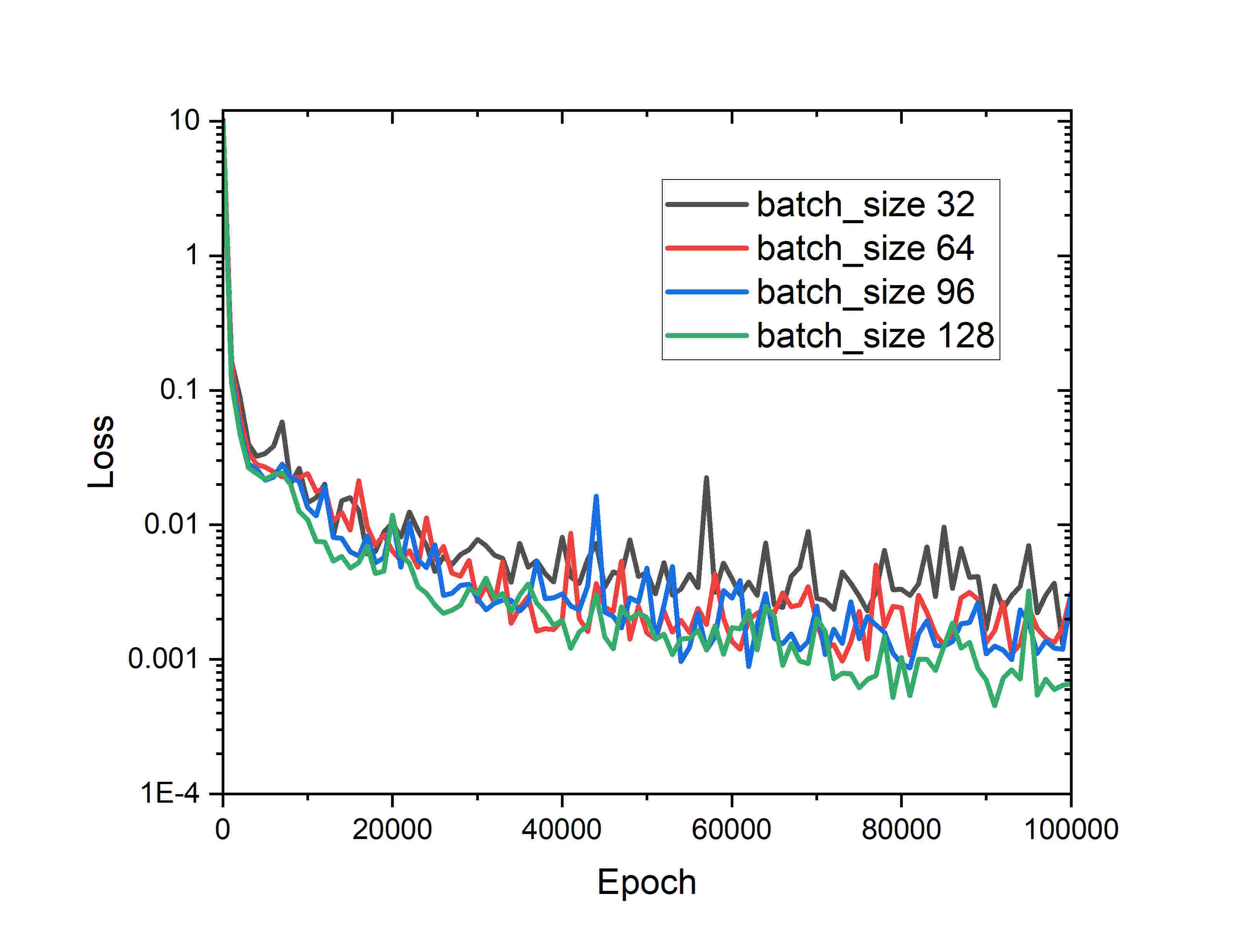}
    \caption{\textsf{Loss function with various batch sizes as a function of epochs for NN2. }}
    \label{fig:hyperpara_bz}
\end{figure*}

\begin{figure*}
    \centering
    \includegraphics[width=4in]{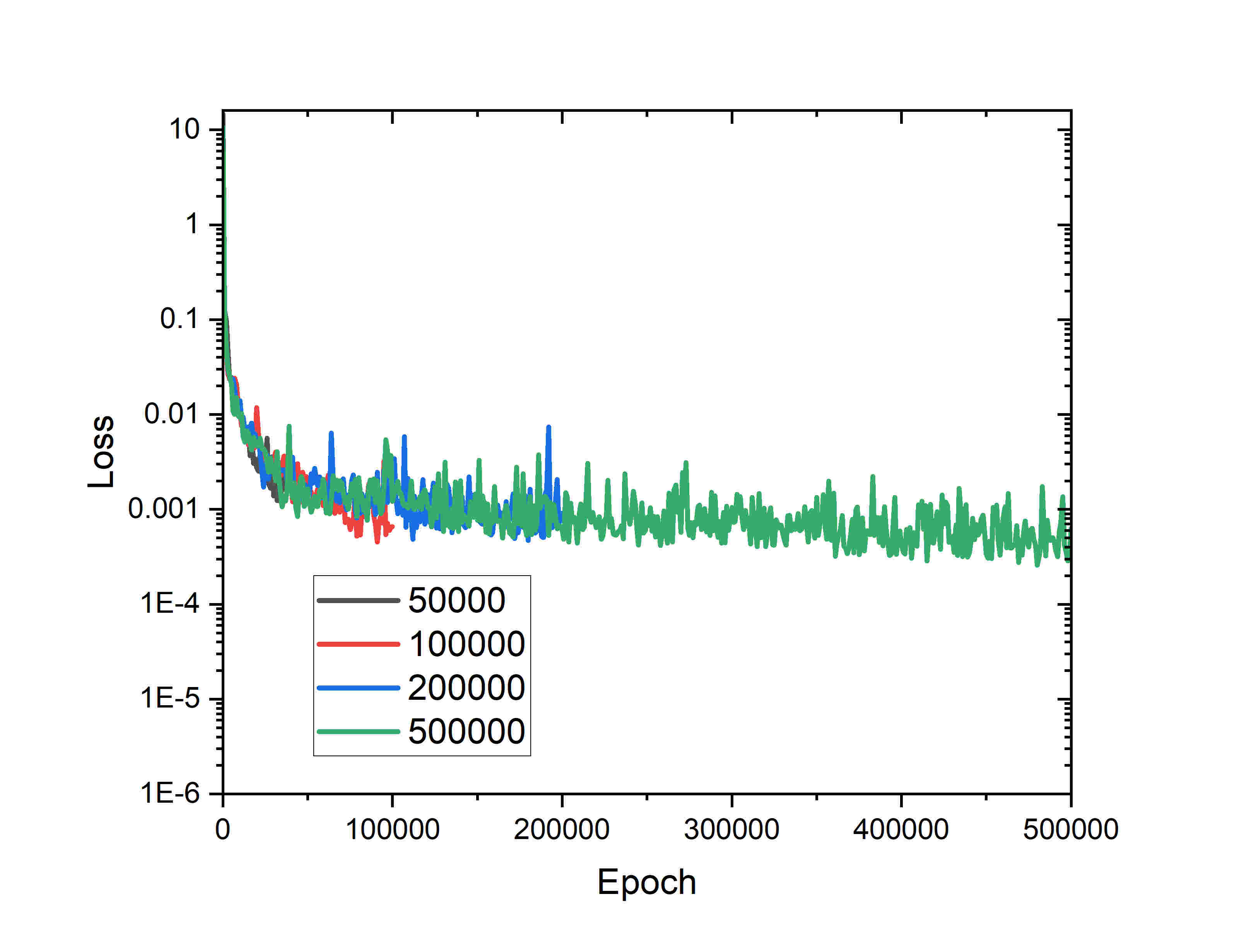}
    \caption{\textsf{Loss function as a function of epochs for NN2. }}
    \label{fig:hyperpara_noe}
\end{figure*}

In this section, we performed CG simulations in a channel using the DPD model. A large constrained fluid model system was built.
The simulation box is 207.4 nm $\times$ 7.764 nm $\times$ 38.83 nm.
A constant acceleration 5$\times$$10^{5}$ \si{m/s^2} was added to the fluid particles to model a Poiseuille flow,
and the velocity data of fluid particles were obtained and subsequently fed to NN2 in our framework. 
The NN2 model was then used to predict the velocity profiles in the system.
The structure of NN2 has been described in Section II. In the NN2 model, there are some hyperparameters that cannot be directly trained from the data or literature.     
The hyperparameters define how our model is actually structured. Because there is no general way to calculate the hyperparameters to reduce the loss in order to find the optimal model architecture, we performed some tests
on the effect of hyperparameters. The learning rate, NN size, number of epochs, activation function, and batch size were tested in this work. Fig. \ref{fig:hyperpara_lr} shows the loss function with various learning rates 
as a function of the number of epochs. The results indicate that 1$\times$$10^{-3}$ is a good candidate for the initial state. And the decay rate should be smaller if an adaptive learning rate is used. The structure of the neural network will affect the prediction 
of the model. Here, we use the structure of the branch net as an example because the structures of the branch net and trunk net are the same.
 Fig. \ref{fig:hyperpara_NNlayer} and \ref{fig:hyperpara_NNwidth} present the result of the neural network structure for the loss function convergence. 
We found that the NN with four hidden layers is large enough for our model. Additional NN layers do not
improve the accuracy of the model. Regarding the NN width, we found that the accuracy showed nonmonotonic changes in the NN width. In our system, when the NN width increases from 60 to 100 neurons, the accuracy will also increase. 
However, the accuracy decreases as the NN width continued to grow from 100 to 120 neurons. The trend is the same as the NN width reaches 160 neurons. That indicates an optimal NN width exists for our model. We also tested several popular
activation functions in our model, including the tanh function, rectified linear activation function (Relu), and sigmoid function. We found that the Relu activation function is the best one for our model, as shown in Fig \ref{fig:hyperpara_af}.
The batch size may influence the dynamics of the learning algorithm and the stability of the neural network model. We tested four batch sizes from 32 to 128. As shown in Fig. \ref{fig:hyperpara_bz}, all the trainings are stable.
The results indicate that the large batch size (128) slightly improves the accuracy in 100,000 epochs compared with the batch size 32 case.
The effect of the number of epochs on the loss function is presented in Fig. \ref{fig:hyperpara_noe}.
Four epochs, i.e., 5$\times$$10^{4}$,  1$\times$$10^{5}$, 2$\times$$10^{5}$, and 5$\times$$10^{5}$, were tested. We found that 5$\times$$10^{4}$ epochs are not enough for the convergence of the loss function in our training.
When the number of epochs is greater than or equal to 1$\times$$10^{5}$, the loss function tends to be flat with increasing epochs. 

\subsection{\label{sec:level31}The second neural network: effect of the boundary condition}
\begin{figure*}
\captionsetup[subfigure]{justification=centering}
\centering
  \begin{subfigure}{0.45\textwidth}
  \includegraphics[width=\linewidth]{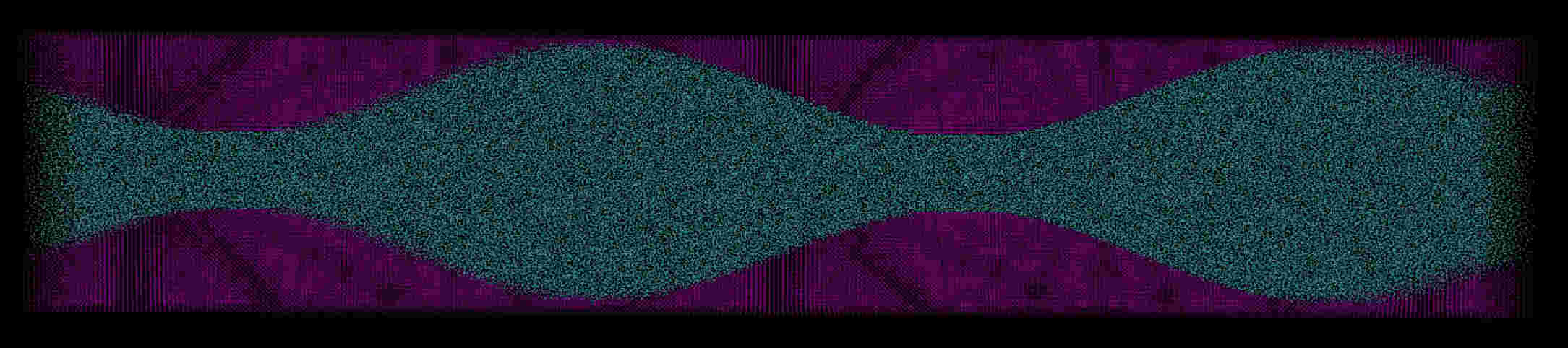}
  \caption{k=0.0125}
  \end{subfigure}
  \begin{subfigure}{0.40\textwidth}
  \includegraphics[width=\linewidth]{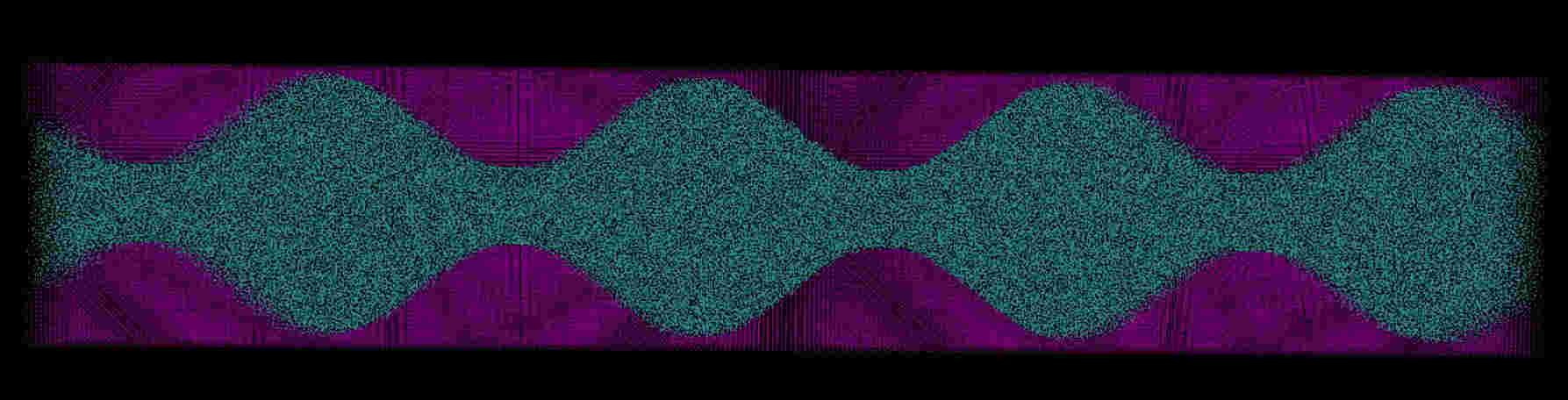}
  \caption{k=0.025}
  \end{subfigure}
  \begin{subfigure}{0.45\textwidth}
  \includegraphics[width=\linewidth]{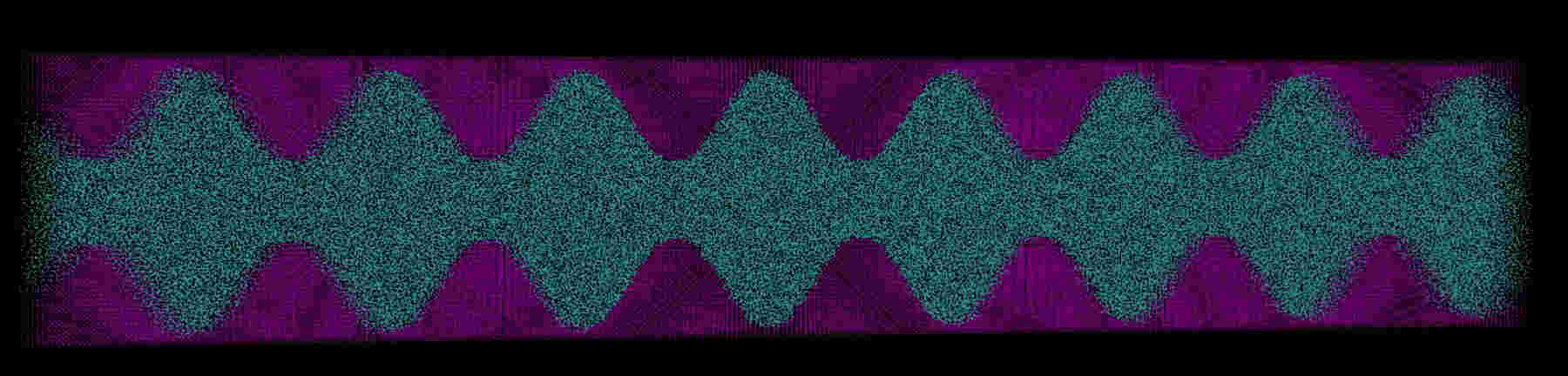}
  \caption{k=0.05}
  \end{subfigure}
  \begin{subfigure}{0.47\textwidth}
  \includegraphics[width=\linewidth]{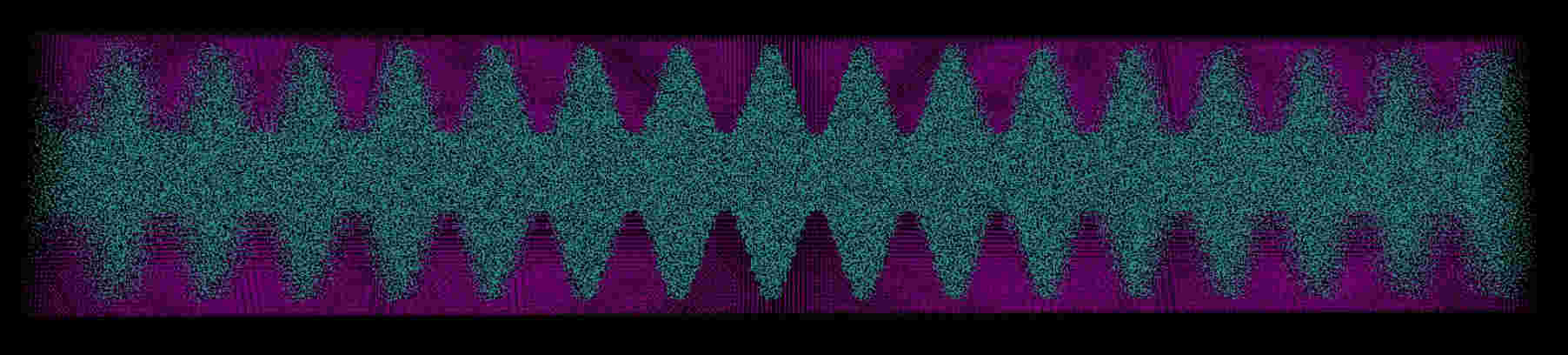}
  \caption{k=0.1}
  \end{subfigure}
  \begin{subfigure}{0.45\textwidth}
  \includegraphics[width=\linewidth]{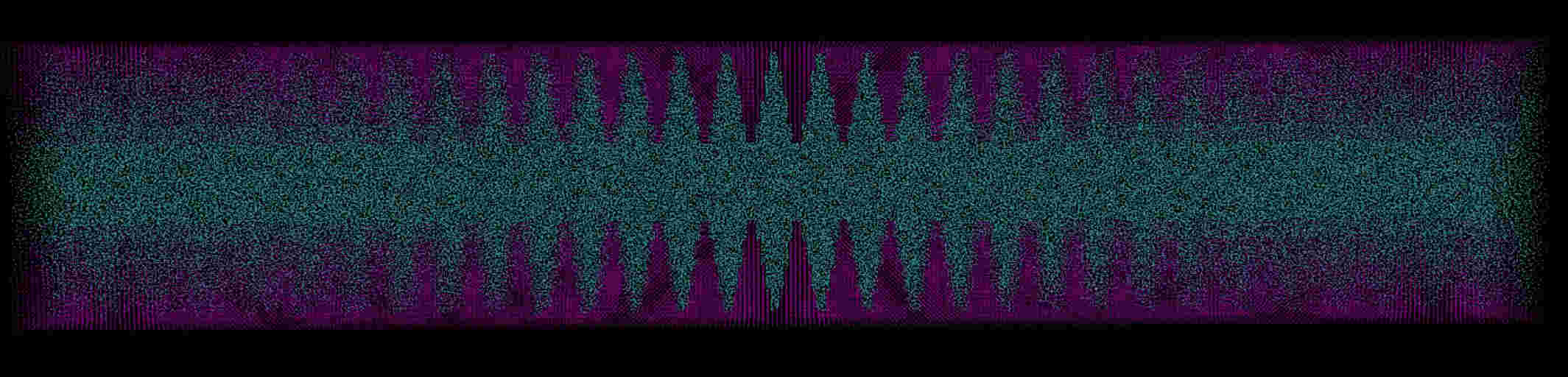}
 \caption{k=0.2}
  \end{subfigure}
    \caption{Snapshots of systems with complex boundary conditions. The center is the fluid and the sides are walls}
    \label{fig:shape}
\end{figure*}

It can be difficult to directly determine the velocity distribution of fluid in a channel with a complex boundary using the laws of physics or classical
methods, especially at the microscopic level. Like petroleum in shale rocks,
the velocity profile of fluid in stream is affected by channel morphology, roughness, and other factors. 
In the above sections, we built a particle-based channel model with complex boundary conditions represented by a sine function to simulate the constrained fluid.
We investigated the effect of complex boundary shape on velocity profile prediction by tuning the period of the sine function.
Five systems with different periods were tested. The snapshots of the systems are shown in Figure \ref{fig:shape}.

\begin{figure*}
    \includegraphics[width=3in]{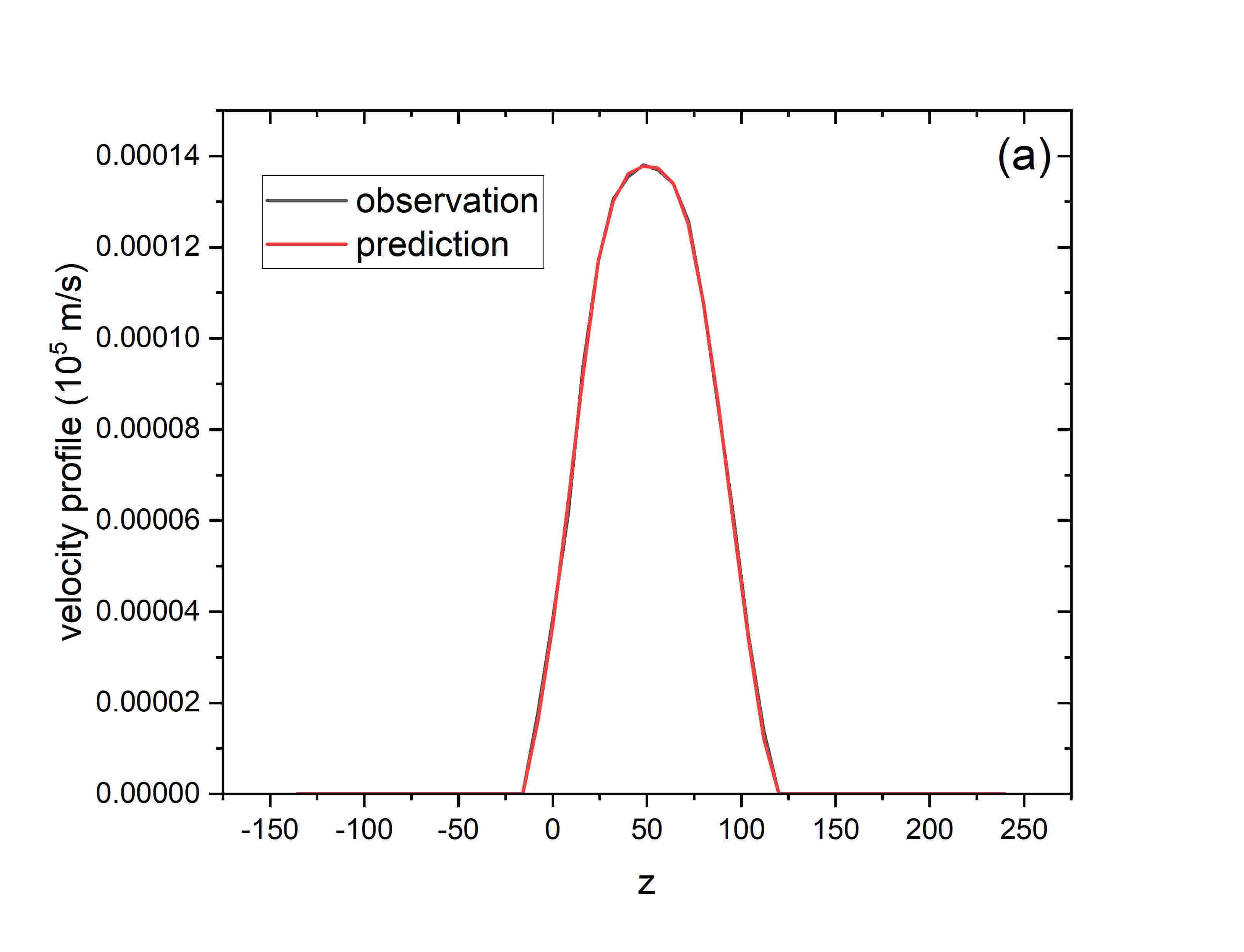}
   \includegraphics[width=3in]{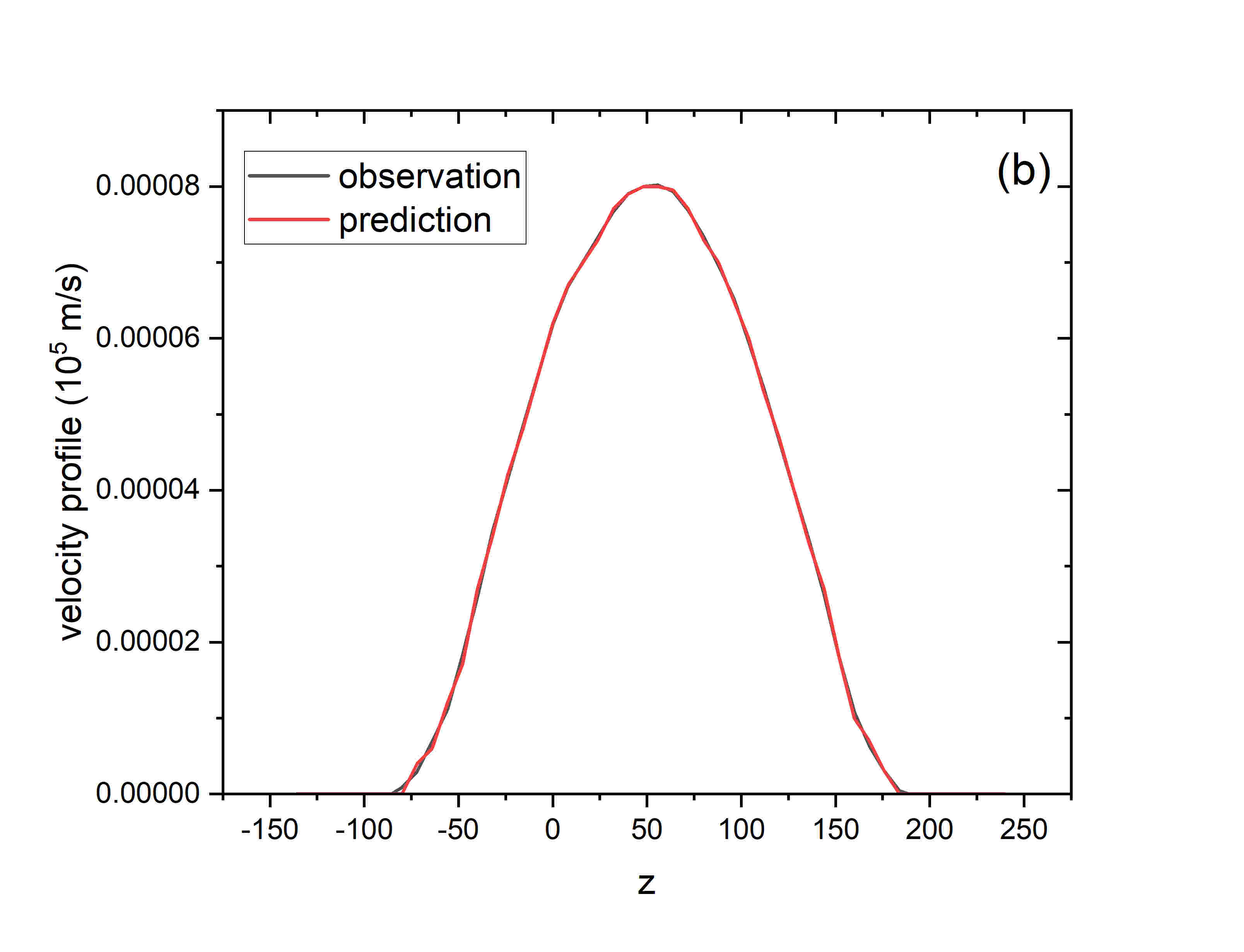}
   \includegraphics[width=3in]{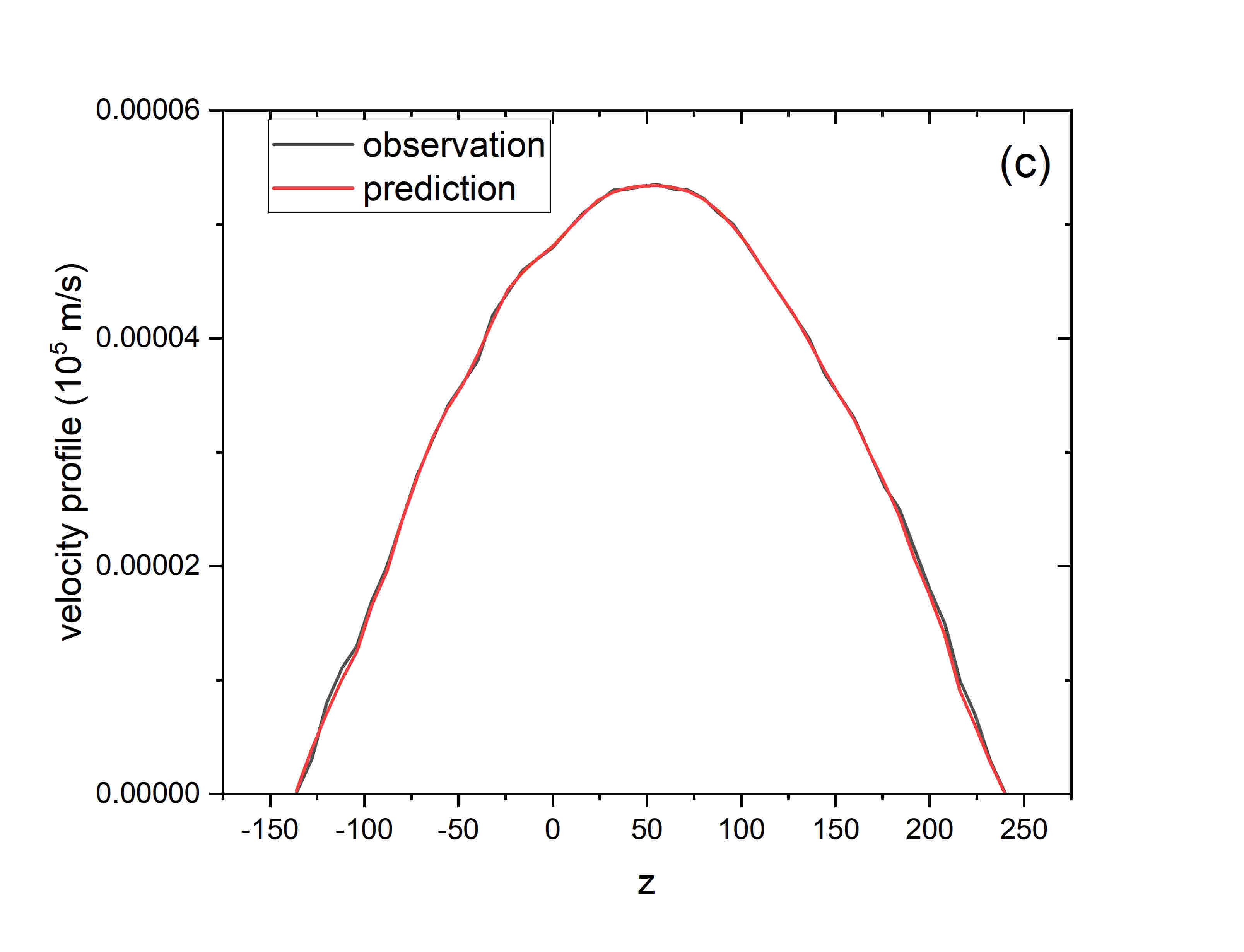}
    \includegraphics[width=3in]{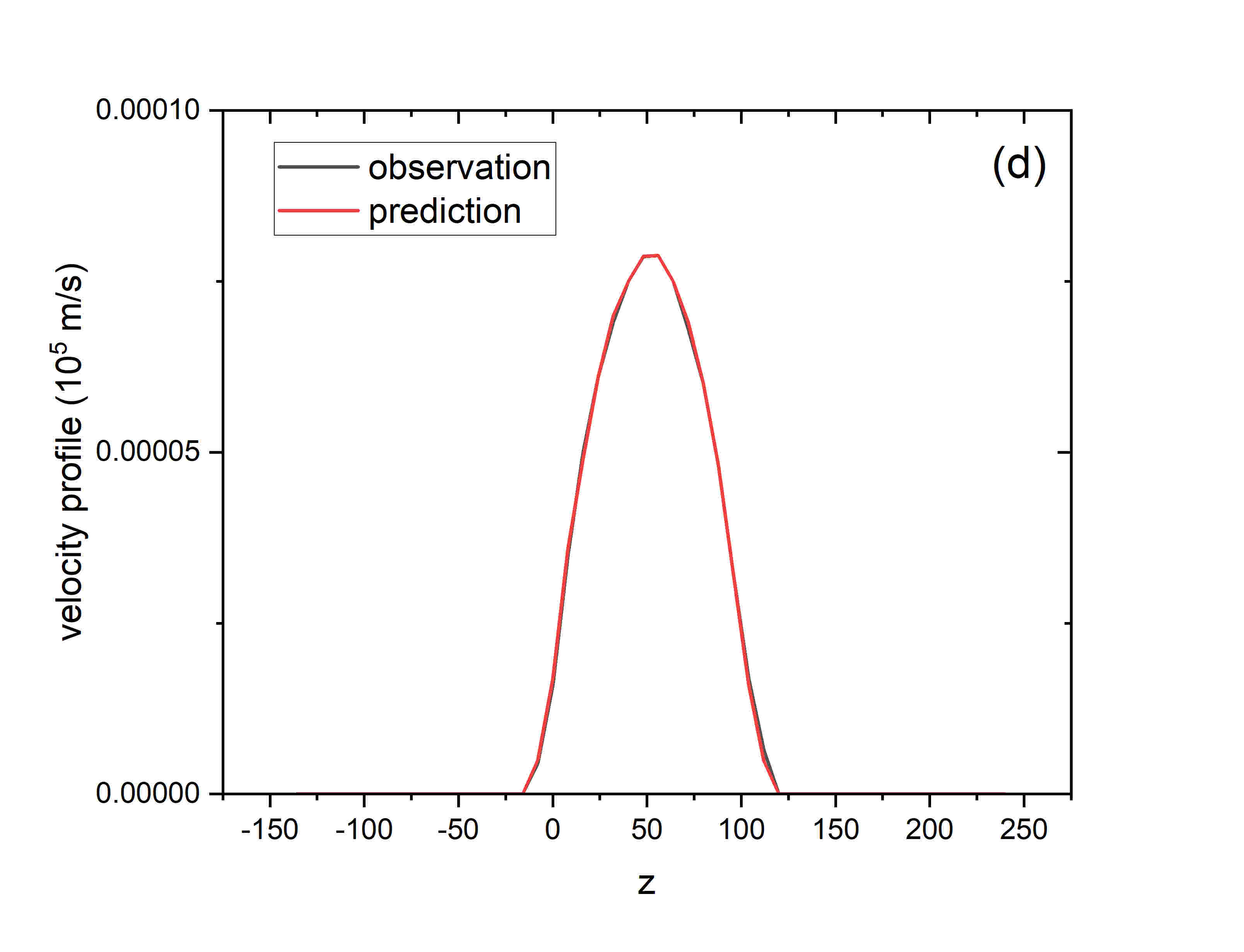}
   \includegraphics[width=3in]{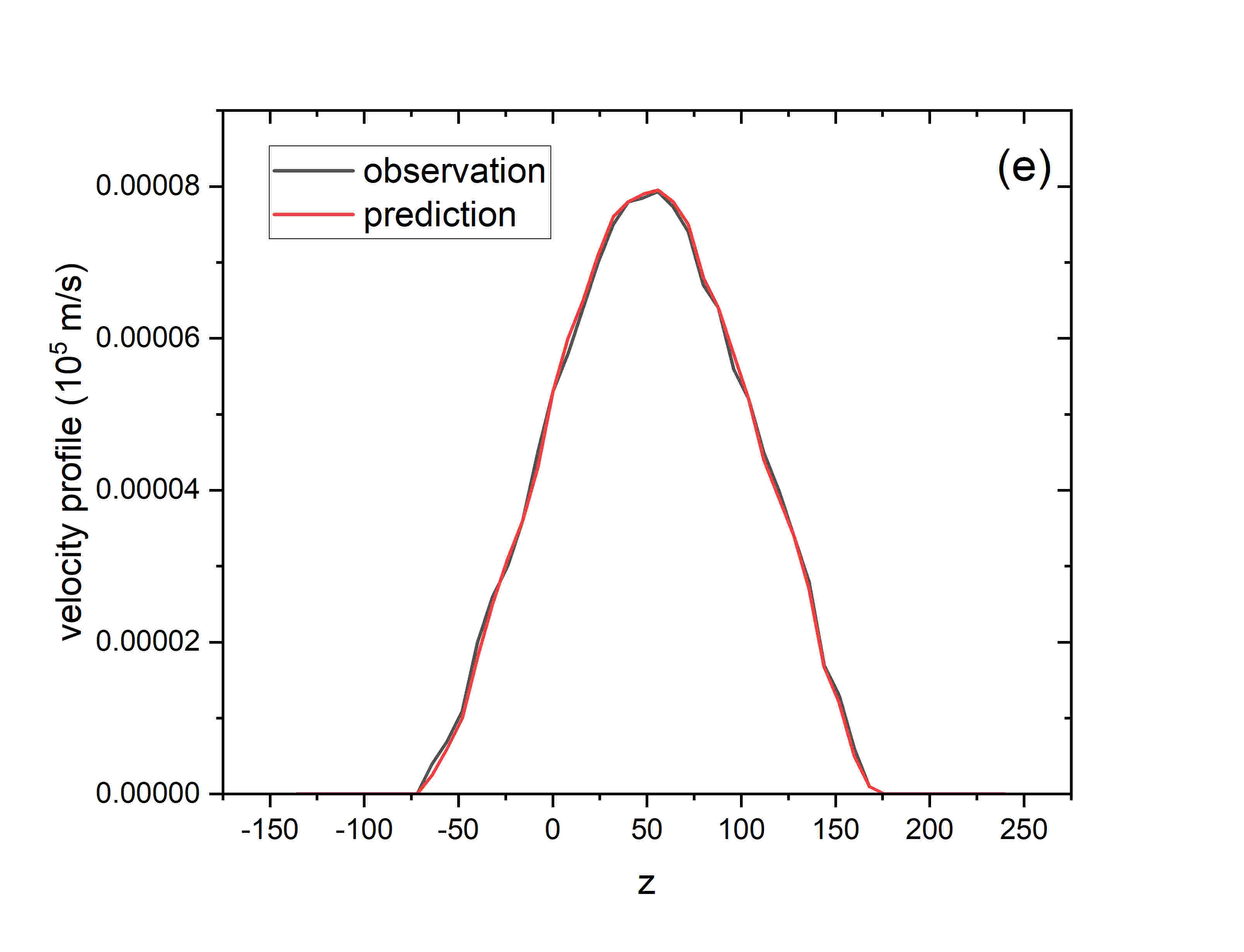}
   \includegraphics[width=3in]{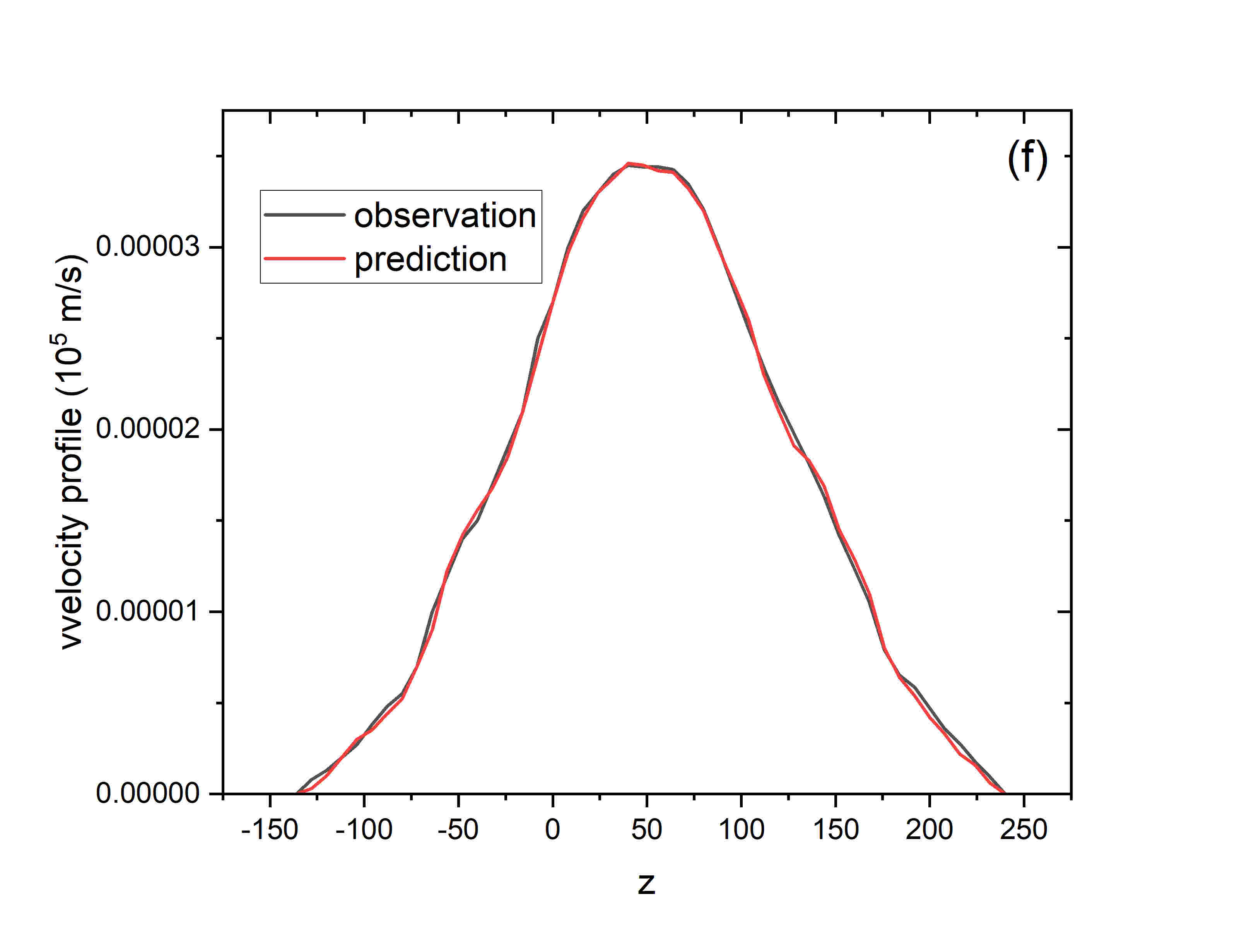}
    \caption{\textsf{The observation and prediction of velocity profiles at different positions for two different boundary conditions: (a-c) at the peak, in the middle, and at the valley for the first boundary condition, and (d-f) at the peak, in the middle, and at the valley
for the second boundary condition}}
    \label{fig:position}
\end{figure*}

Figure \ref{fig:position} shows the velocity profiles obtained in the CG simulation at different positions in the system. 
The average velocity of fluid will be slower with a more complex boundary shape.
It was reported that the structured boundary surfaces may induce ordering on the plane on which they bound the fluid, but disordering is found
in the perpendicular direction in the equilibrium DPD simulation where the structured boundary surface is very rough.\cite{Terron-Mejia2015}
This phenomenon induced an order–disorder phase transition on a mesoscopic scale. While it is only observed when the particle size is larger than the distance between the two neighbored peaks of the sine-function shape boundary.\cite{Terron-Mejia2015}
In our simulation, we can see that the particles exist in the region between the two neighbored peaks of the sine-function shape boundary, even for the smallest one. 
That means the particle size is less than the distance between the two neighbored peaks of the sine-function shape boundary. Therefore, the order–disorder phase transition is not present in our simulations.  
The velocity profile was predicted by the second neural network with a DeepONet architecture. 
Interestingly, we found that the predictions of the NN model are not very good at certain positions for some cases, e.g., $k$ = 0.2 in our systems.
Some positions that are close to the boundary cannot be effectively predicted by the NN model. These were found in the systems with a large curvature on the boundary.
In systems with smooth boundary conditions, however, this was not observed. This may be because of the large curvature of the boundary and the lack of valid data in these regions.
On one hand, the particle may be trapped locally in the simulation when the boundary curvature is large, which would reduce the efficiency of sampling in the system during the data collection simulation.
On the other hand, the large curvature of the boundary may discretize the velocity profile, posing challenges in representing it accurately using a continuous function with a NN.
When $k$ is very large, i.e., the particles are unable to come into the region between the two neighboring peaks of the sine-function shape boundary in simulation, the velocity data quality may be improved as well as the prediction result compared with the rough surfaces because the detectable curvature of the boundary is not large in this case. While this is beyond the current work and needs further investigation in the future. 
For these available cases, the effect of position on model prediction accuracy is further investigated. Figure \ref{fig:position} shows the predicted velocity profiles of different positions for the center bump. As mentioned above,
the boundary condition is described by a sine function. We only change the coefficient $k$. 
The data in Figure \ref{fig:position} a-c were obtained 
from the system with sine function boundary condition $k$ = 0.0125, and data in Figure \ref{fig:position} d-f were obtained 
from the system with sine function boundary condition $k$ = 0.025. 
Generally speaking, the NN model can effectively predict the velocity profile in these cases. The relative errors of Figure \ref{fig:position} a, b, and c are 2.0\%, 3.1\%, and 4.4\%, respectively.
For both cases, the prediction at the position of the peak of the boundary is a little better than of the middle and valley positions.
This may be because the position of the peak is close to the bulk region of the fluid and the "curvature" at the peak position is small in simulations. Also, we can  anticipate relatively improved prediction for the first system because the boundary of the first system is smoother than that of the second case. 
The relative errors of Figure \ref{fig:position} d, e, and f are 3.5\%, 4.4\%, and 5.9\%, respectively. This result validates our assumption. 
More training data may further reduce the prediction differences between the peak and valley and between different boundary conditions, while the cost increases as more physical simulations are required.

\section{\label{sec:level4}Conclusions and future work}
In this work, a multiscale modeling framework for constrained fluid with ML technique was built. The thermodynamic, structural, and dynamic properties of the fluid constrained by solid surfaces with complex shapes were studied in
detail. The atomic-scale and mesoscale models were connected by integrating double NNs. 
The octane liquid was selected as an example. The first NN represents the CG potential that connected the atomic model and the CG model. The CG model with NN potential can reproduce the structural and thermodynamic properties of octane in an atomistic simulation at the CG level.
The first NN model was trained in a bottom-up way. 
The NN CG model was integrated to a DPD framework to improve the dynamics of the model.
By combing with the DPD framework, we investigated the fluid dynamics of the constrained octane fluid with complex boundary shapes through physical simulation and built a surrogate model with the second NN. 
The velocity data and the boundary condition parameters were fed to the second NN with a DeepONet architecture. The velocity profiles of the constrained octane fluid can be predicted using the second NN. 
It is possible to apply this to any other constrained fluids and connect the current model with a macroscopic model via the velocity field, pressure field, and so on. 
We investigated the effect of boundary shape with the prediction of NN. The velocity of the fluid will be slower with a more complex boundary shape, and this is consistent with the roughness concept at the macroscopic level.
We identified that the NN model can effectively predict the velocity profile generally. However, the prediction is not as good when the local curvature of the boundary is quite large.
We found that the prediction at the position near the bulk fluid region is somewhat better than the position that is close to the solid boundary.
This may be because of the spatial inhomogeneity of the sampling efficiency for the dataset.

We demonstrated that with the aid of a NN, the CG model trained with thermodynamic properties can also effectively reproduce the structural property in an atomic simulation.
This has been observed in previous bottom-up CG models.\cite{Izvekov2005, Zhang2018} As the Henderson theorem indicates\cite{HENDERSON1974}, the solution of a CG model with two-body interaction potential is not unique when matching the structural property of atomistic and CG models such as RDF.
With the top-down CG method, the NN method also provides a way to find the optimum potential parameters in a two-body interaction potential or a many-body interaction potential framework. It may need additional physical or chemical properties at the macroscopic level as constraints,
 although the many-body interaction that is implicitly included in the NN potential could improve the reproducibility of the potential energy surface in a CG model.
 One method is differentiating through the numerical solution of Newton’s equations of motion
to obtain gradients \cite{Greener2021,Thaler2021,wang2023} . This method would be very efficient for phase space sampling, and more data can be generated to improve the existed interaction potentials, which can be combined into our future work to build the CG potential in a hybrid way. 
Overall, this work provides some insights into multiscale model development with ML techniques. 
The approach will be beneficial to micro/nanofluidic applications for enhanced recovery of oil from complex pores and electrolyte transport in batteries.

\section*{\label{sec:level5}Conflicts of interest}
There are no conflicts to declare.

\section*{\label{sec:level6}Acknowledgments}
This work is supported by the U.S. Department of Energy, Office of Science, Advanced Scientific Computing Research program
under the Collaboratory on Physics-Informed Learning Machines for Multiscale and Multiphysics Problems (PhILMs) project (Project No. 71268), and under the Scalable, Efficient and Accelerated Causal Reasoning Operators, Graphs and Spikes for Earth and Embedded Systems (SEA-CROGS) project (Project No. 80278).
 Pacific Northwest National Laboratory is a multi-program national
 laboratory operated for the U.S. Department of Energy by Battelle Memorial Institute
 under Contract No. DE-AC05-76RL01830.

\nocite{*}
\section*{\label{sec:level7}References}
\bibliography{aipsamp}

\end{document}